\documentclass[%
 twocolumn,
groupedaddress,
 amsmath,amssymb,
 aps,
]{revtex4-2}

\usepackage{graphicx}
\usepackage{dcolumn}
\usepackage{bm}
\usepackage{xcolor}
\usepackage[flushleft]{threeparttable}
\usepackage[colorlinks,linkcolor=blue,anchorcolor=blue,citecolor=blue,]{hyperref}
\newcolumntype{.}{D{.}{.}{-1}}
\newcolumntype{d}[1]{D{.}{.}{#1}}
\newcommand*{\wn}{cm$^{-1}$}
\newcommand*{\hsm}{H$_{2}$S}
\newcommand*{\Hm}{H$_{2}$}
\newcommand*{\X}{X$^1\Sigma_g^+$}
\newcommand*{\EF}{EF$^1\Sigma_g^+$}
\newcommand*{\F}{F$^1\Sigma_g^+$}

\renewcommand{\eqref}[1]{Eq.~(\ref{#1})}

\begin{document}

\title{Precision measurement of quasi-bound resonances in H$_2$ and the H + H scattering length} \thanks{Tribute to the late Prof. Lutoslaw Wolniewicz and honoring him for his inspirational contributions over many decades to the theory of the hydrogen molecule.}

\author{K.-F. Lai}
 \affiliation{Department of Physics and Astronomy, LaserLaB, Vrije Universiteit \\
 De Boelelaan 1081, 1081 HV Amsterdam, The Netherlands}
 \author{E. J. Salumbides}
 \affiliation{Department of Physics and Astronomy, LaserLaB, Vrije Universiteit \\
 De Boelelaan 1081, 1081 HV Amsterdam, The Netherlands}

 \author{M. Beyer}%
 \affiliation{Department of Physics and Astronomy, LaserLaB, Vrije Universiteit \\
 De Boelelaan 1081, 1081 HV Amsterdam, The Netherlands}

  \author{W. Ubachs}%
  \affiliation{Department of Physics and Astronomy, LaserLaB, Vrije Universiteit \\
 De Boelelaan 1081, 1081 HV Amsterdam, The Netherlands}

\date{\today}

\begin{abstract}
Quasi-bound resonances of H$_2$ are produced via two-photon photolysis of H$_2$S molecules as reactive intermediates or transition states, and detected before decay of the parent molecule into three separate atoms.
As was previously reported [K.F. Lai et al., Phys. Rev. Lett. 127, 183001 (2021)] four centrifugally bound quantum resonances with lifetimes of multiple $\mu$s, lying energetically above the dissociation limit of the electronic ground state X$^1\Sigma_g^+$ of H$_2$, were observed as X($v,J$) = (7,21)$^*$, (8,19)$^*$, (9,17)$^*$, and (10,15)$^*$,
while also the short-lived ($\sim 1.5$ ns) quasi-bound resonance X(11,13)$^*$ was probed.
The present paper gives a detailed account on the identification of the quasi-bound or shape resonances, based on laser detection via  \F-\X\ two-photon transitions, and their strongly enhanced Franck-Condon factors due to the shifting of the wave function density to large internuclear separation.
In addition, the assignment of the rotational quantum number is verified by subsequent multi-step laser excitation into autoionization continuum resonances.
Existing frameworks of full-fledged ab initio computations for the bound region in H$_2$, including Born-Oppenheimer, adiabatic, non-adiabatic, relativistic and quantum-electrodynamic contributions, are extended into the energetic range above the dissociation energy.
These comprehensive calculations are compared to the accurate measurements of energies of quasi-bound resonances, finding excellent agreement.
They show that the quasi-bound states are in particular sensitive to non-adiabatic contributions to the potential energy.
From the potential energy curve and the correction terms, now tested at high accuracy over a wide range of energies and internuclear separations, the s-wave scattering length for singlet H+H scattering is determined at $a = 0.2735^{39}_{31}~a_0$.
It is for the first time that such an accurate value for a scattering length is determined based on fully ab initio methods including effects of adiabatic, non-adiabatic, relativistic and QED with contributions up to $m\alpha^6$.
\end{abstract}

\maketitle

\section{Introduction}

Chemically bonded diatomic molecules in their ground electronic configuration typically support large but finite numbers of quantum levels, which are assigned with vibrational and rotational quantum numbers ($v,J$). These bound levels lie energetically below the dissociation threshold. Advanced calculations including non-adiabatic, relativistic and quantum-electrodynamical (QED) contributions for H$_2$ reveal that this simplest neutral molecule has 302 such bound states in its ground electronic configuration (X$^1\Sigma_g^+$)~\cite{Komasa2011}.
All these bound rovibrational states exhibit long lifetimes in excess of $10^5$ s~\cite{Black1976}.
The computational methods for the determination of H$_2$ level energies witnessed great progress, starting with the wave-mechanical explanation that H$_2$ is a bound system by Heitler and London \cite{Heitler1927}.
This evolved further to quantum-mechanical methods for the calculation of the dissociation energy of H$_2$, using two-centered wave functions by James and Coolidge \cite{James1933} and the computations by Kolos and Roothaan~\cite{Kolos1960b}, via the accurate computations including relativistic effects by Wolniewicz \cite{Wolniewicz1993,Wolniewicz1995}, to the state-of-the-art approaches involving 4-particle variational calculations augmented with QED computations \cite{Simmen2013, Puchalski2018,Wang2018,Puchalski2019,Puchalski2019b}.

Progress on the theoretical side was matched by increasingly precise measurements of the dissociation energy of the smallest neutral molecule \cite{Herzberg1961,Herzberg1969,Balakrishnan1992,Zhang2004,Liu2009,Cheng2018,Beyer2019,Holsch2019}, now finding agreement between theory and experiment at the level of $10^{-5}$ \wn.
The computations were verified via comparison with measurements of high-rotational angular momentum states~\cite{Salumbides2011} and vibrational splittings in H$_2$~\cite{Cheng2012,Hu2012,Campargue2012,Dickenson2013,Niu2014}.
These improvements both on the experimental and theoretical side have made the hydrogen molecule into a test ground for investigating the effect of fifth forces~\cite{Salumbides2013}, extra dimensions of space-time~\cite{Salumbides2015b}, and for probing physics beyond the Standard Model~\cite{Ubachs2016}.

In addition to the large set of bound quantum states there exists a class of molecular resonances bound by a positive centrifugal barrier of size $\hbar^2 J(J+1)/2\mu R^2$ (with $\mu$ the reduced mass of the molecule and $R$ the internuclear distance) added to the potential energy of the rotationless molecule. Such resonances lie energetically above the dissociation threshold giving rise to quasi-bound states that are prone to tunneling through this centrifugal barrier, a process known as rotational predissociation \cite{HerzbergBook}. Such features were discussed in terms of a 'mechanical instability' of molecules after observation of the breaking off of a rotational progression in the ground state of the HgH molecule as early as 1929, at the dawn of molecular quantum spectroscopy  \cite{Oldenberg1929}. In that study it was explained that there exist only a limited number of such centrifugally bound states; for increasing values of $J$ the potential energy again becomes repulsive. A similar phenomenon was observed in the case of AlH \cite{Bengtsson1930}.
Recently, the quasi-bound resonances in the hydrogen molecular ion  were probed and accurately measured via multi-step laser spectroscopy, revealing the X$^+\,^2\Sigma_g^+$($v^+, J^+$) = (18,4)$^*$ resonance in para-H$_2^+$, and the X$^+$(17,7)$^*$ resonance in ortho-H$_2^+$ \cite{Beyer2016,Beyer2016b}. These resonances in the three-particle quantum system appeared to have lifetimes of 10 - 25 ps due to their rapid tunneling through the centrifugal barrier.

The quasi-bound states studied here have previously been observed and assigned by Dabrowski~\cite{Dabrowski1984} and by Roncin and Launay~\cite{Roncin1994} in emission spectra of the Lyman and Werner bands with optical transitions terminating on such resonances.
Over the years computations were performed for such resonances in the hydrogen neutral molecule \cite{Leroy1971,Selg2011,Selg2012} and these indicate that some of the resonances are long-lived with lifetimes in excess of $\mu$s, and even up to few 100 $\mu$s. This makes these quasi-bound resonances amenable in precision experiments with lasers of narrow bandwidth.



The ultraviolet photolysis of \hsm\  has proven to be an ideal target system for the production of highly energetic rovibrational states in the electronic ground configuration of \Hm.
In an explorative study, Steadman and Baer demonstrated the production of vibrationally-excited H$_2$ as a product channel from \hsm\ photodissocation \cite{Steadman1989}.
They assigned transitions between high-vibrational states in H$_2$ to both inner and outer well states in the \EF\ double well potential.
However, in the controlled two and three-laser studies probing X($v''=11-14$) \cite{Niu2015b,Trivikram2016,Trivikram2019,Lai2021} only transitions to the \F\ outer well could be confirmed.
In an alternative study, starting from the photolysis of H$_2$CO, therewith producing H$_2$ vibrational states of intermediate quantum numbers X($v''=3-9$), two-photon transitions to both \EF\ inner and outer well were found~\cite{Quinn2020}.
Precision spectroscopic measurements probing the vibrational states just below the dissociation threshold allowed for tests of molecular \emph{ab initio} calculations for these quantum states exhibiting wave function density at large internuclear separation~\cite{Lai2021}.

The present study, of which a preliminary report was published previously~\cite{Lai2021b}, explores the energy range just above the dissociation threshold of H$_2$ upon two-photon UV-photolysis of hydrogen sulfide molecules.
This results in the detection of quasi-bound states in H$_2$, using 2+1' resonance-enhanced multi-photon ionization, and in the accurate determination of the excitation energies of the long-lived resonances. State-of-the-art quantum calculations developed for the bound region are extended to the region above threshold and compared with measured resonance energies.
Hence, the experiments on quasi-bound resonances allow for testing the potential energy curve of the H$_2$ molecule for a wide range of internuclear separations, that had not been explored before.
The well-tested H$_2$ potential may then be applied to re-examine the H+H scattering dynamics and for the first time determine a value of the scattering length including effects of adiabatic, non-adiabatic, relativistic and QED with contributions up to $m\alpha^6$.
Via this way the precision measurement of the quasi-bound resonances provides detailed quantitative information on atomic scattering that uniquely bridges the gap between the ultracold physics of atomic hydrogen towards the chemical formation of molecular hydrogen.

\section{Experimental}

\begin{figure}[ht]
\begin{center}
\includegraphics[width=\linewidth]{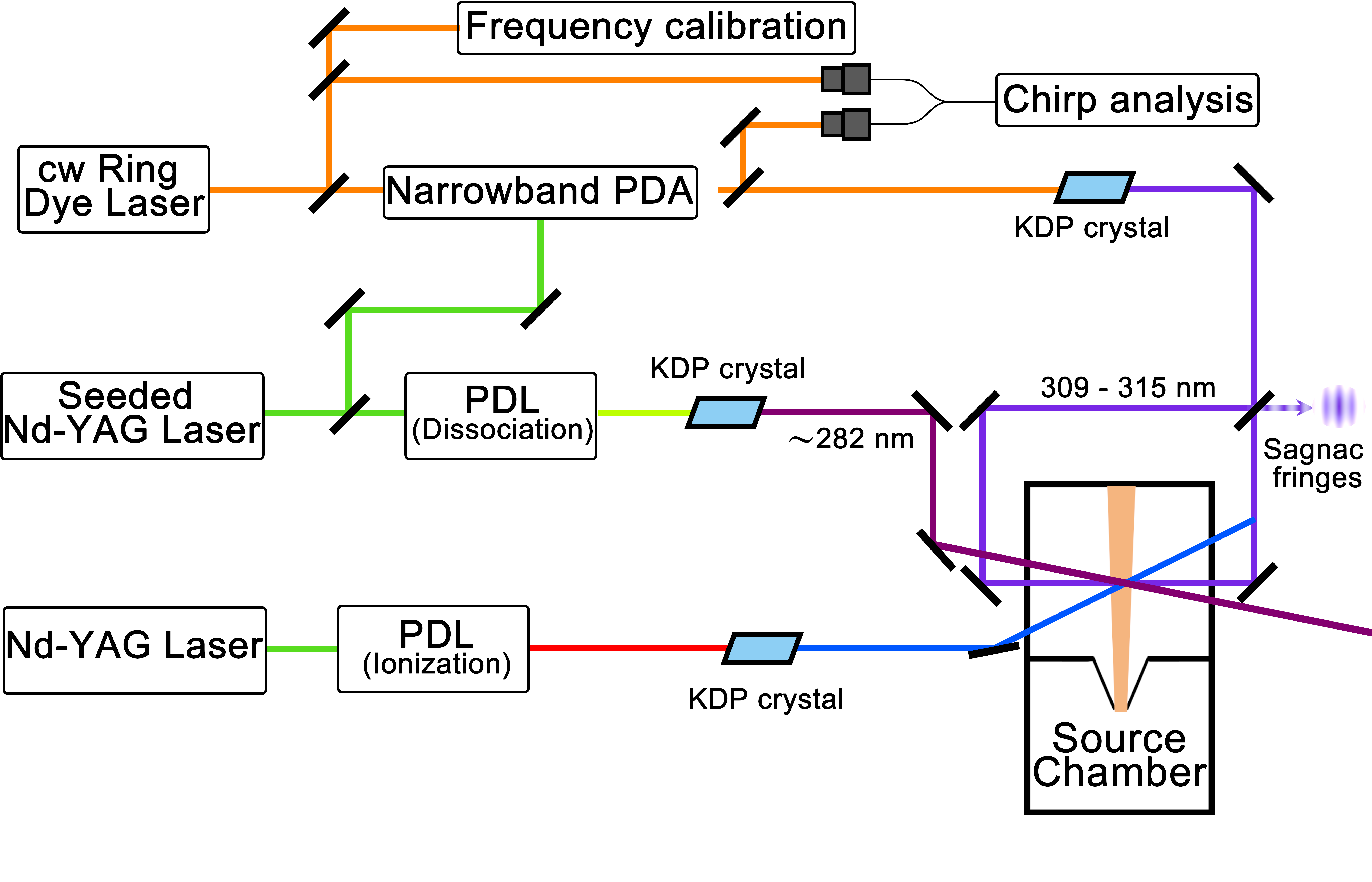}
\caption{\label{setup}
Layout of the experimental setup involving three independently tunable UV-lasers, and a molecular beam apparatus. For details see text.
}
\end{center}
\end{figure}

\begin{figure}[b]
\begin{center}
\includegraphics[width=0.95\linewidth]{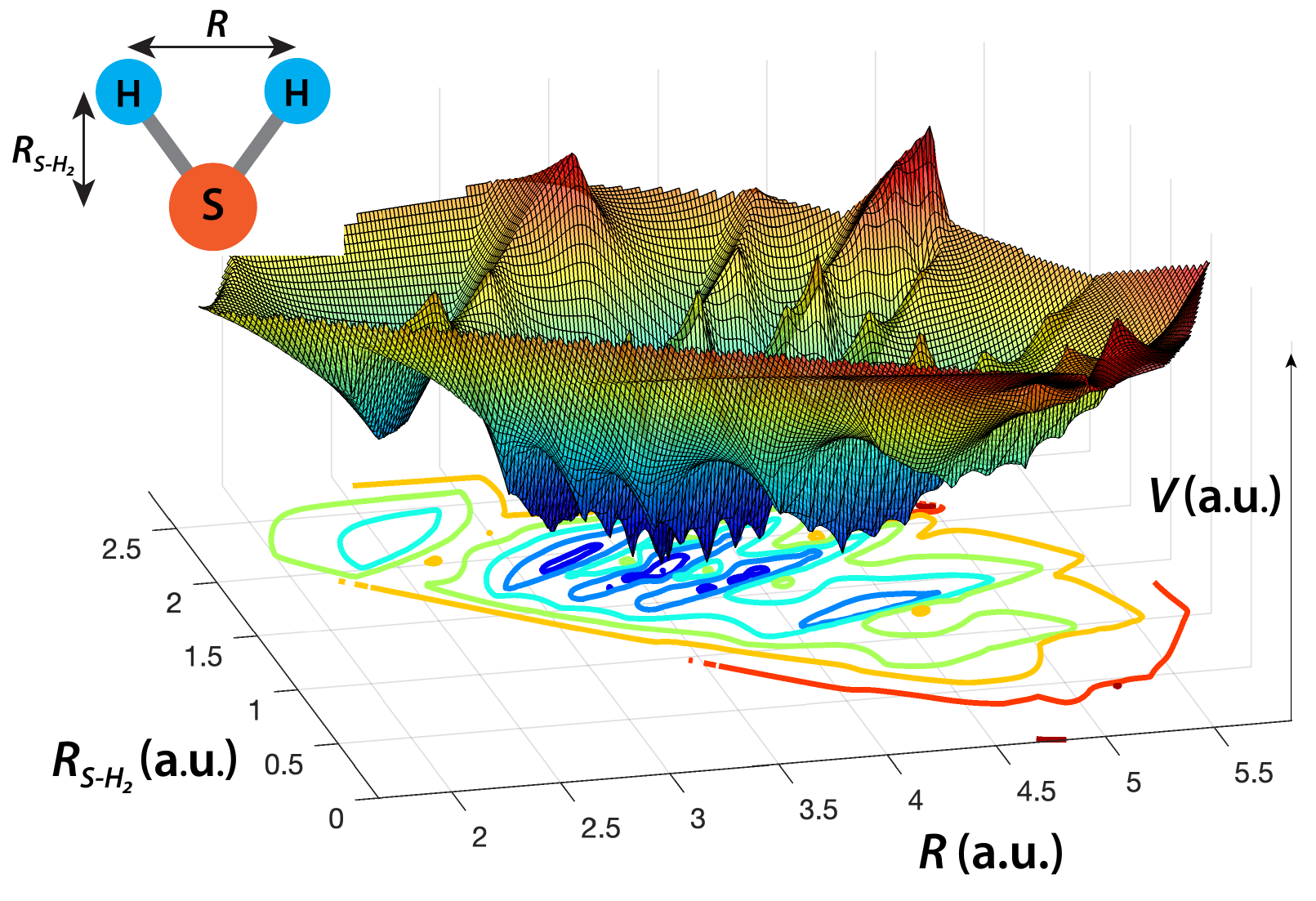}
\caption{\label{fig:photolysis}
Potential energy surface for the electronic ground states of H$_2$S \cite{Tarczay2001} as a function of the H-H ($R$) and S-H$_2$ distance.
}
\end{center}
\end{figure}

The experimental setup, displayed in some detail in Fig.~\ref{setup}, is essentially similar to the one used previously for the detection of bound resonances below the dissociation threshold \cite{Trivikram2016,Trivikram2019,Lai2021}.
Two-photon UV-photolysis proceeds via the path:
\begin{equation*}
    \text{H$_{2}$S} \xrightarrow{2h\nu} \text{S} (^{1}{\rm D}_{2}) + \text{H$_{2}^{*}$}
   \rightarrow \text{S} (^{1}{\rm D}_{2}) + \text{H}(1s) + \text{H}(1s),
\end{equation*}
where H$_2^*$ signifies a quasi-bound resonance existing as an intermediate reaction product before tunneling and disintegrating in two H($1s$) atoms.  To illustrate the efficiency of this excitation scheme regarding the production of the shape resonances, the potential energy surface of the electronic ground state of H$_2$S is displayed in Fig.~\ref{fig:photolysis}, as a function of the S-H$_2$ and H-H distance. The minimum of the potential is located around $R_{\rm S-H_2}\approx 1.7~a_0$ and $R_{\rm H-H}\approx3.6~a_0$. Comparing this with the potential energy curve of the electronic ground state of H$_2$ (see lower part of Fig.~\ref{fig:photolysis}), it can be seen that the second classical turning point is located at a similar value of $R_{\rm H-H}$, with the corresponding maximal amplitude of the radial wave function. This minute change of the proton-proton distance explains the efficient production of the shape resonances, in line with the Franck-Condon principle.

\begin{figure}[b]
\begin{center}
\includegraphics[width=0.95\linewidth]{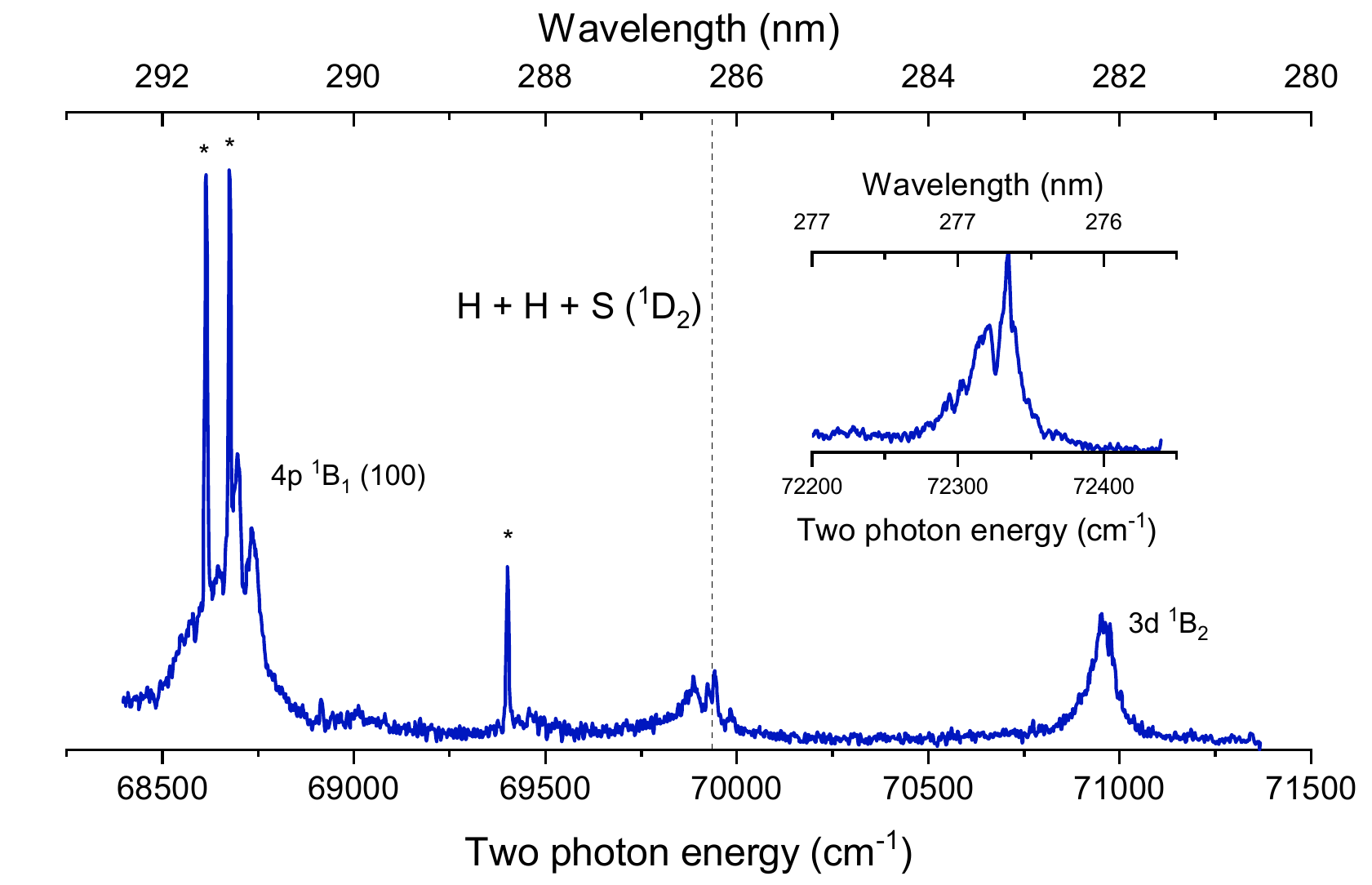}
\caption{\label{fig:H2S-excitation}
Two-photon excitation spectrum of the  of H$_2$S molecule with probing of H$_2$S$^{+}$ ion species. The three narrow signals labeled with asterisk corresponding to two-photon transitions in the sulphur atom. See text for details.
}
\end{center}
\end{figure}

An excitation spectrum of the \hsm\ molecule, probed via 2+1 UV resonance-enhanced multi-photon ionization is shown in Fig ~\ref{fig:H2S-excitation}.
Two relatively strong and broad resonances, at excitation wavelengths between 276 and 292 nm, are observed in the H$_2$S$^+$ ion channel, and are therefore associated with excited states in \hsm.
The sharp resonances, labeled with asterisks, are identified as strong S$^+$ ion signals from two-photon excitation of the S-atom, ($^4$S) $6p$ $^5$P$_1$, ($^2$D) $4p$ $^1$P$_1$, ($^2$D) $4p$ $^3$D$_1$ and ($^2$D) $4p$ $^1$F$_3$ - 3$p^4$ $^1$D$_2$~\cite{Steadman1988b}. Production  and precision spectroscopy of S atoms resulting from \hsm\ photolysis was investigated previously in our setup under similar conditions~\cite{Lai2020c}.
The strongest two resonances peaking at excitation energies of 68700 \wn\ and 70900 \wn\ are assigned as two-photon excitation to $4p$ $^1$B$_1$ (100) and $3d$ $^1$B$_2$, respectively.
Processes related to the first level have been  discussed extensively by Steadman et al. \cite{Steadman1988b}.
The latter level is based on assignments in the absorption spectrum of Masuko et al. \cite{Masuko1979}.
The selection rule for two-photon transitions from the $^1$A$_1$ ground state in \hsm\ allows to observe $^1$A$_1$, $^1$A$_2$, $^1$B$_1$ and $^1$B$_2$, which cover all possible one-photon transitions.
Masuko et al. proposed that the broad features peaking at wavelengths of 140.7 nm belong to two overlapping transitions, to $3d\,^1$A$_1$ and $3d\,^1$B$_2$ at 140.73 nm and 140.90 nm, respectively.
However, Mayhew et al. assigned this feature to $3d\,^1$B$_2$ with a maximum at 140.72 nm based on the pattern of $n = 4$ in the same Rydberg series~\cite{Mayhew1987}.
The ab initio calculations on \hsm\ excited state level energies and oscillator strengths have shown that the higher energy $3d\,^1$A$_1$ state has a stronger one-photon absorption compared to $3d\,^1$B$_2$~\cite{Pericoucayere1997}, in agreement with the assignment in Ref.~\cite{Masuko1979}.
The resonance peaking at 281.8 nm in the present work, equivalent to 140.9 nm in one-photon transition, shows a nearly symmetric profile, in contrast with the doublet structure observed in the absorption spectrum.
Based on the limited information available, we follow the assignment of Ref.~\cite{Masuko1979}.
In addition, a pair of weak doublet structures observed at excitation at wavelengths of 286.0 and 276.5 nm are interpreted as part of a vibrational progression $3d\,^1$B$_1$ (000) and (100). Further study is needed to support the assignments.

The photolysis resonance at 291 nm was used for the production and investigation of H$_2$ molecules in $v=11$ vibrational states~\cite{Trivikram2019}.
For the production of H$_2$ in $v=13,14$ levels photolysis on the two-photon resonance at 281.8 nm was employed~\cite{Lai2021}.
In the present experiment the wavelength of the dissociation laser is tuned to this two-photon absorption resonance in \hsm\ at 281.8~nm, providing sufficient energy to overcome the dissociation energy, at 69935(25)~\wn, for a complete three particle dissociation~\cite{Zhou2020,Zhao2021} with an excess of 1000 \wn\ above threshold.
Here it is noted that by spin selection rules, photolysis of H$_2$S only produces H$_2$ molecules in combination with sulphur atoms in the $^{1}{\rm D}_{2}$ electronically excited state.
During all measurements this first photolysis laser is fixed at this wavelength of 281.8 nm, at the peak of the two-photon dissociation resonance of H$_2$S, for yielding an optimal amount of H$_2^*$.
The excess energy is then released as kinetic energy in the product, giving H$_2^*$ an additional momentum which makes them rapidly escape from the interaction zone.
A fly-out time of 20 ns was measured from the interaction zone of the three overlapping UV-lasers.

\begin{figure}
\begin{center}
\includegraphics[width=0.9\linewidth]{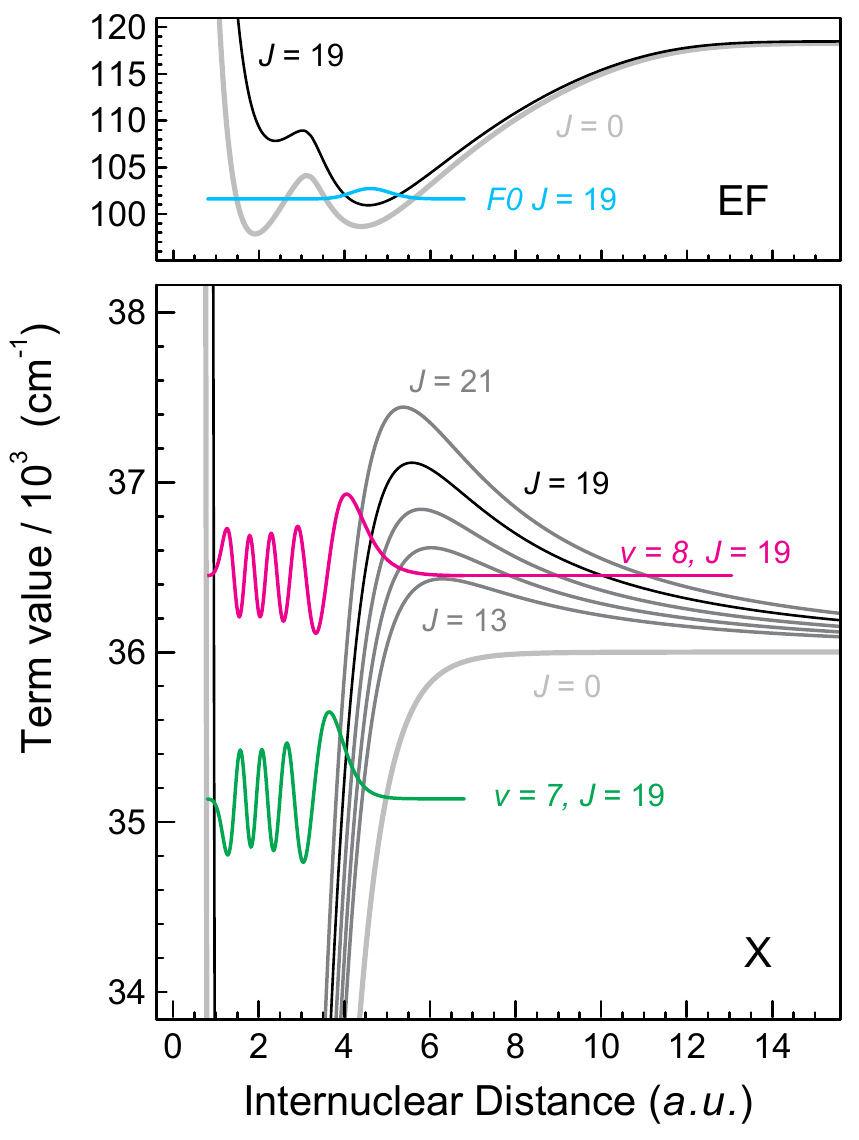}
\caption{\label{ex_scheme}
Potential energy curves for the \X\ ground state and the \EF\ excited states, drawn for the $J=0$ rotationless energies and for higher odd $J$ values, including the centrifugal energy. Also wave functions for a bound X(7,19) and quasi-bound X(8,19)$^*$ level are displayed. The wave function for the F0($J=19$) excited level (upper panel) indicates overlap at large internuclear separation (4-5 a.u.). }
\end{center}
\end{figure}

The H$_2^*$ short-lived resonances are probed in a three laser scheme, where two-photon UV photolysis of \hsm\ is followed by excitation of H$_2^*$ via two-photon Doppler-free excitation into the  \F, $v = 0$ electronically excited outer well state, denoted as F0 in the following.
In the potential energy diagram of the H$_2$ molecule, depicted in Fig.~\ref{ex_scheme}, the wave functions of bound and a quasi-bound levels are plotted as a function of internuclear separation and at their respective excitation energies. It illustrates the binding and tunneling of these states, while it also shows the favorable Franck-Condon overlap of their wave functions with the F0 level in the excited outer well.
A third UV laser then further excites the F0 population into the autoionization continuum, preferably on a strong autoionization resonance, after which H$_2^+$ species can be detected for signal recording.

The frequency calibration of the spectroscopy laser, probing the excitation of quasi-bound resonances to the F0-outer well state, is crucial because it determines the accuracy at which the energy of those resonances are determined.
This laser is a traveling-wave Pulsed-Dye-Amplifier (PDA) amplifying the output of a continuous-wave (CW) ring-dye-laser, upon-frequency doubling delivering a pulsed output with a frequency bandwidth of $\sim 100$ MHz~\cite{Eikema1997}.
The absolute frequency calibration relies on saturated absorption spectroscopy of hyperfine-resolved I$_2$ using the CW-output of the ring laser as well as a wavemeter (Toptica High-Finesse WSU-30)~\cite{Lai2020}.
Effects of frequency chirp in the pulsed output of the PDA is analysed and corrected for, following established methods~\cite{Eikema1997}.
Excitation of the two-photon transitions is established in a Doppler-free geometry with counter-propagating beams aligned in a Sagnac interferometric scheme~\cite{Hannemann2007}.

The pulse sequences of the  UV-lasers, all three with pulse durations  $\sim  5$ ns, are such that delays of at least 10 ns are imposed to avoid possible AC-Stark effects induced by the photolysis and ionization lasers and to obtain the highest accuracy on the determination of level energies of the quasi-bound resonances.
The AC-Stark effects of the spectroscopy laser are systematically investigated with extrapolation to zero-power densities.
These procedures lead to measurement of
two-photon transition frequencies  upon averaging over multiple measurement sequences.
The resulting accuracies of the X-EF spectroscopic measurements are discussed in Section~\ref{detect}.

For most of the measurements the ion optics are triggered at least 80~ns delayed from the spectroscopy laser, therewith creating  a DC-field-free environment.
For the weak transitions prompt H$_2^+$ ions produced from \hsm\ photolysis are separated by applying a DC-field of 1.3 kV/cm of opposite polarity during excitation.
It was experimentally verified that such DC-fields do not cause a Stark shift at the present precision level.
This finding is in agreement with calculations based on polarizabilities for the EF state~\cite{Komasa2005} and X state~\cite{Kolos1967}, predicting a less than 1 MHz shift for the DC-field used.

\section{Theory}
\label{theory}

In support of the experimental studies several calculations are performed. For testing the level structure, and the combination differences between the \X($v,J$)$^*$ quasi-bound resonances, highly accurate computations are carried out for these levels. For assigning the quantum numbers of the quasi-bound resonances by their \F\ - \X\ transitions, computations are performed of the F0 levels, at a lower level of precision.

\subsection{The \X\ state}
\label{sec:theoryX}
Level energies for the quasi-bound states are calculated using the nonrelativistic quantum electrodynamics (NRQED) approach, in which relativistic, leading-order radiative and higher order QED corrections are added to a nonrelativistic Hamiltonian.
The nonrelativistic energy is traditionally evaluated within the Born-Oppenheimer (BO) approximation by keeping the position of the nuclei fixed.
Because of the light mass of the nuclei with respect to the mass of the electron, the adiabatic correction, scaling with $m_\text{e}/m_\text{p}$, and the nonadiabatic correction to the BO approximation are substantial in case of the hydrogen molecule.
Nonrelativistic level energies can also be obtained directly by a variational solution of the full four-body Schr\"odinger equation, without invoking the BO approximation \cite{Simmen2013,Wang2018,Puchalski2018,Puchalski2019}. Such beyond-BO calculations allow to reach an accuracy limited by the uncertainties of the fundamental constants, but are computationally expensive, because a separate variational calculation is needed for every single ro-vibrational level.

Nonadiabatic perturbation theory (NAPT)~\cite{Czachorowski2018}  maintains the computational efficiency of the BO approach through separating the electronic and nuclear Schr\"odinger equation, while preserving accuracy.
This is achieved by including $R$-dependent corrections to the potential energy curve and by employing $R$-dependent reduced masses in the nuclear Schr\"odinger equation, accounting for leading-order nonadiabatic interactions on the order $(m_\text{e}/m_\text{p})^2$.
These methods were previously explored by Kutzelnigg and coworker~\cite{Kutzelnigg2007,Jacquet2008}.
Because no full four-body nonrelativistic energies are available for the quasi-bound states reported here, level energies were obtained using NAPT as presented in Ref.~\cite{Komasa2019}.

The radial nuclear Schr\"odinger equation (in atomic units) within NAPT is given by \cite{Komasa2019}
\begin{align}\label{eq:SEraw}
    \left[-\frac{1}{R^{2}} \frac{\partial}{\partial R} \frac{R^{2}}{2 \mu_{\|}(R)} \frac{\partial}{\partial R}+\frac{J(J+1)}{2 \mu_{\perp}(R) R^{2}}+\mathcal{V}(R)\right] \phi_{i}(R) \notag\\
=E_i \phi_{i}(R),
\end{align}
with the $R$-dependent vibrational ($\mu_{\|}$) and rotational ($\mu_{\perp}$) reduced masses
\begin{align}
    \frac{1}{2\mu_{\|}(R)} & = \frac{1}{2\mu_\text{a}}+ W_{\|}(R), \text{and} \\
    \frac{1}{2\mu_{\perp}(R)} & = \frac{1}{2\mu_\text{a}}+ W_{\perp}(R),
\end{align}
with $\mu_\text{a}=(m_\text{p}+1)/2$ being the reduced atomic mass. The functions $W_{\|}(R)$ and $W_{\perp}(R)$ are defined in Ref.~\cite{Komasa2019} and vanish for $R\to\infty$. The potential $\mathcal{V}(R)$ is given by
\begin{equation}\label{eq:potterms}
  \mathcal{V}(R) = \mathcal{E}_\text{BO} + \mathcal{E}_\text{ad} + \delta\mathcal{E}_\text{na},
\end{equation}
representing the adiabatic \cite{Pachucki2014} and nonadiabatic \cite{Pachucki2015} corrections to the BO potential energy curve \cite{Pachucki2010}.

Commonly, a new radial function is defined as $f_i(R)=R\phi_i(R)$ to remove terms involving the first derivative $\partial/\partial R$ in \eqref{eq:SEraw}, leading to a simplified equation involving only the second derivative and some multiplicative factors. However, because of the $R$-dependence of $\mu_{\|}(R)$, a term $W_{\|}'(R)(\partial/\partial R)$ remains, where $W_{\|}'(R)$ indicates the first derivative of $W_{\|}(R)$ with respect to $R$. Using the ansatz $\chi_i(R)=R\phi_i(R)\exp{(-Z(R))}$ in \eqref{eq:SEraw}~\cite{Leroy1985,Selg2011}, the first derivative term vanishes for $Z(R)$ fulfilling
\begin{equation}
    \frac{d\,Z(R)}{d\,R} = \mu_{\|} W_{\|}'(R).
\end{equation}
The radial Schr\"odinger equation, Eq.~(\ref{eq:SEraw}), can then be written as:
\begin{align}\label{eq:SE_transf}
\left[-\frac{1}{2\mu_{\|}(R)} \frac{d^2}{d\,R^2} - \frac{\mu_{\|}(R)}{2} \left(W_{\|}'(R) \right)^2 + \frac{1}{2}W_{\|}''(R) \right.\notag\\
\left.+ \frac{1}{R}W_{\|}'(R) + \frac{J(J+1)}{2 \mu_{\perp}(R) R^{2}}+\mathcal{V}(R)\right] \chi_{i}(R) =E_i \chi_{i}(R),
\end{align}
which can be solved using the renormalized Numerov method introduced by Johnson~\cite{Johnson1977}.

Relativistic ($m\alpha^4$) and QED ($m\alpha^5$, $m\alpha^6$) corrections were evaluated as described in Ref.~\cite{Komasa2019} using the BO nuclear wavefunction, obtained by solving \eqref{eq:SE_transf} using the nuclear reduced mass $\mu_n=m_p/2$, $\mathcal{V}(R)=\mathcal{E}_\text{BO}$ and $Z(R)=0$. The nonadiabatic part of the relativistic correction is known to be important and was also included \cite{Komasa2019}, whereas we ignored higher order QED and finite size corrections, which are below 1~MHz for the states considered here.

The potential energy functions, relativistic and QED corrections, as well as the $R$-dependent reduced masses and its derivatives were interpolated on a grid with $0.001~a_0$ stepsize in the range $0.1a_0$ to $R_\text{max}=25a_0$ using the H2SPECTRE program \cite{Komasa2019}. The CODATA2018 \cite{CODATA2018} values were used for the fundamental constants: $E_h=219474.63136320(43)~\text{cm}^{-1}$, $m_p=1836.15267343(11)$ and $\alpha=7.2973525693(11)\times 10^{-3}$.

When comparing the nonadiabatic level energies, relativistic and QED corrections of the bound states obtained by solving \eqref{eq:SE_transf} with the renormalized Numerov method, we found exact agreement with the results obtained by using H2SPECTRE, which is based on a discrete-variable representation~\cite{Colbert1992}.
The current version of H2SPECTRE does however not allow the calculation of quasi-bound states and resonances.

As for the quasi-bound states, resonance positions and widths are determined by calculating the phase shift for a given $J$ by propagating the wave function to large internuclear distance, where
\begin{equation}
    \lim_{R\to\infty} Z(R) = 0
\end{equation}
and
\begin{equation}\label{CRH:eq:WF_bessel}
\lim_{R\to\infty}\chi(R;k) \propto kR(j_{J}(kR)\cos\eta_{J} - n_{J}(kR)\sin\eta_{J}) ,
\end{equation}
where $j_{J}$ and $n_{J}$ are the spherical Bessel functions and $k=\sqrt{2\mu_\text{a}(E-\mathcal{V}(\infty))}$.
The phase shift for a given energy $\eta_{J}(E)$ was derived from the values of the wave function at the two outermost grid points $R_\text{a}$ and $R_\text{b}=R_\text{max}$ using
\begin{equation}\label{CRH:eq:phase_shift_tan}
\tan{\eta_{J}} = \frac{ Kj_{J}(kR_\text{a}) - j_{J}(kR_\text{b}) }{ Kn_{J}(kR_\text{a}) - n_{J}(kR_\text{b}) }; ~ K=\frac{ R_\text{a}\chi_{J}(R_\text{b}) }{ R_\text{b}\chi_{J}(R_\text{a})  }.
\end{equation}
The energy grid in the vicinity of a resonance was made adaptive by requiring a certain number of points per phase jump $\pi$.

\begin{table*}
\renewcommand{\arraystretch}{1.3}
\caption{Contributions to the dissociation energies of the observed shape resonances of H$_2$ in cm$^{-1}$. A negative dissociation energy indicates that these levels lie above the dissociation threshold. A comparison is made with results from Selg~\cite{Selg2012}.
\label{tab:calc_res}}
\begin{tabular}{l.....}
Contribution / X$(v,J)^*$ & \multicolumn{1}{c}{(11,13)$^*$} & \multicolumn{1}{c}{(10,15)$^*$}   & \multicolumn{1}{c}{(9,17)$^*$}  & \multicolumn{1}{c}{(8,19)$^*$} & \multicolumn{1}{c}{(7,21)$^*$}  \\
\hline
$m\alpha^2$ & -192.2683(29) &-186.1598(36)  & -224.5890(40) & -327.0262(43) & -505.4818(45) \\
\quad adiabatic     & 3.1895 & 4.2857    & 5.3013    & 6.2827    & 7.2555 \\
\quad nonadiabatic  & 2.6847 & 3.2666    & 3.6583    & 3.9268    & 4.1120  \\
$m\alpha^4$         & -0.2254 & -0.2945   & -0.3540   & -0.4076   & -0.4570 \\
\quad BO            & -0.2261 & -0.2952 & -0.3546  & -0.4081   & -0.4575 \\
\quad nonadiabatic  & 0.0007 & 0.0006  & 0.0006   &  0.0005  & 0.0004 \\
$m\alpha^5$         & -0.0008 & 0.0002    & 0.0021    & 0.0047    & 0.0079 \\
$m\alpha^6$         & -0.0001 & -0.0001   & -0.0001   & -0.0001   & -0.0001 \\[0.5em]
Total       & -192.4945(29) & -186.4542(36)     & -224.9410(40)     & -327.4291(43) & -505.9310(45) \\
Comparison \cite{Selg2012}  & -192.50 & -186.46 & -224.95 & -327.43  & -505.93 \\
\hline
$\Gamma_\text{FWHM}$ (kHz)  & \multicolumn{1}{c}{$1\times10^{5}$} & 70      &  1    &  0.5     &  1\\
Lifetime (s) & \multicolumn{1}{c}{$2\times10^{-9}$} & \multicolumn{1}{c}{$2\times10^{-6}$}      &  \multicolumn{1}{c}{$2\times10^{-4}$}     &  \multicolumn{1}{c}{$3\times10^{-4}$} & \multicolumn{1}{c}{$2\times10^{-4}$}\\
\hline
\end{tabular}
\end{table*}

Resonance parameters are determined within the collision-time-delay approach developed by Smith \cite{Smith1960a}. For a single channel, the scattering matrix is $S=\exp\left[ 2i\eta_{J} \right]$ and the time-delay matrix $Q$ is given by
\begin{equation}\label{eq:scatQ}
Q=-iS^*\frac{dS}{dE}= 2 \frac{d\eta_{J}}{dE} =  \frac{\varGamma}{(E-E_\text{res})^2+(\varGamma/2)^2},
\end{equation}
where the energy-dependence of the background phase shift has been neglected.
The resonance position corresponds to the position of the maximum of $Q$, i.e., the energy at which $\frac{d\eta_{J}}{dE}$ is maximal (point of inflexion of the phase-shift curve). The level width is given by
\begin{equation}\label{eq:Qgamma}
\varGamma^\text{(Q)} = \frac{4}{Q(E_\text{res})}= \frac{2}{\frac{d\eta_{J}}{dE}\big|_{E=E_\text{res}}}.
\end{equation}

The quasi-bound states in ortho-H$_2$ are known to have narrow width \cite{Selg2012} and a very fine energy grid would be required to locate an increase of $\pi$ in the phase shift. To obtain a first estimate of the level positions we choose to extend the effective potential in \eqref{eq:potterms}: the maximum height of the centrifugal barrier $\mathcal{V}_\text{bar}$ at $R_\text{bar}$ is determined and we set $\mathcal{V}(R>R_\text{ext})=0.5\cdot\mathcal{V}_\text{bar}$ for $R_\text{ext}$ being determined when $\mathcal{V}(R_\text{ext}>R_\text{bar})=0.5\cdot\mathcal{V}_\text{bar}$. Under these conditions the quasi-bound states will appear as bound states and can be easily located.
Nonadiabatic level energies found using the modified potential agree within 4~MHz for X(11,13)$^*$ and to better than 200~kHz for X(7,21)$^*$, X(8,19)$^*$, X(9,17)$^*$ and X(10,15)$^*$ with the energies found using \eqref{eq:scatQ}.
The nonadiabatic level energies for the five resonances in ortho-H$_2$ experimentally observed are given in Table~\ref{tab:calc_res}.

The relativistic and QED corrections summarized in Table~\ref{tab:calc_res} are calculated using the BO nuclear wave functions obtained using the modified potential, which allows to circumvent the use of energy-normalized continuum wave function in the NAPT approach. This is expected to be an excellent approximation for the aimed precision and was verified by a scattering calculation that included the relativistic and QED corrections in \eqref{eq:potterms}. Deviations of level positions between 10 and 22~MHz were found, which we attribute to the fact that the nonadiabatic contributions to the relativistic correction ($\sim20$~MHz) are taken into account at a different level of approximation.
The full-width-at-half-maximum (FWHM) is obtained using \eqref{eq:Qgamma} and is given together with the natural lifetime at the bottom of Table~\ref{tab:calc_res}.

The theoretical uncertainties are determined by the leading order terms neglected in the NAPT approach, which are estimated by scaling the second-order nonadiabatic $m\alpha^2$ corrections with $1/\mu_n$ \cite{Komasa2019} and are found to be between 87 and 135~MHz for the quasi-bound states.
The computation of binding energies for the H$_2^*$ quasi-bound resonances by Selg~\cite{Selg2012}, at a claimed accuracy of 0.01 \wn, are found to agree with the present values  within 0.01 \wn\ (or 300 MHz).

\begin{figure*}
\begin{center}
\includegraphics[width=0.63\linewidth]{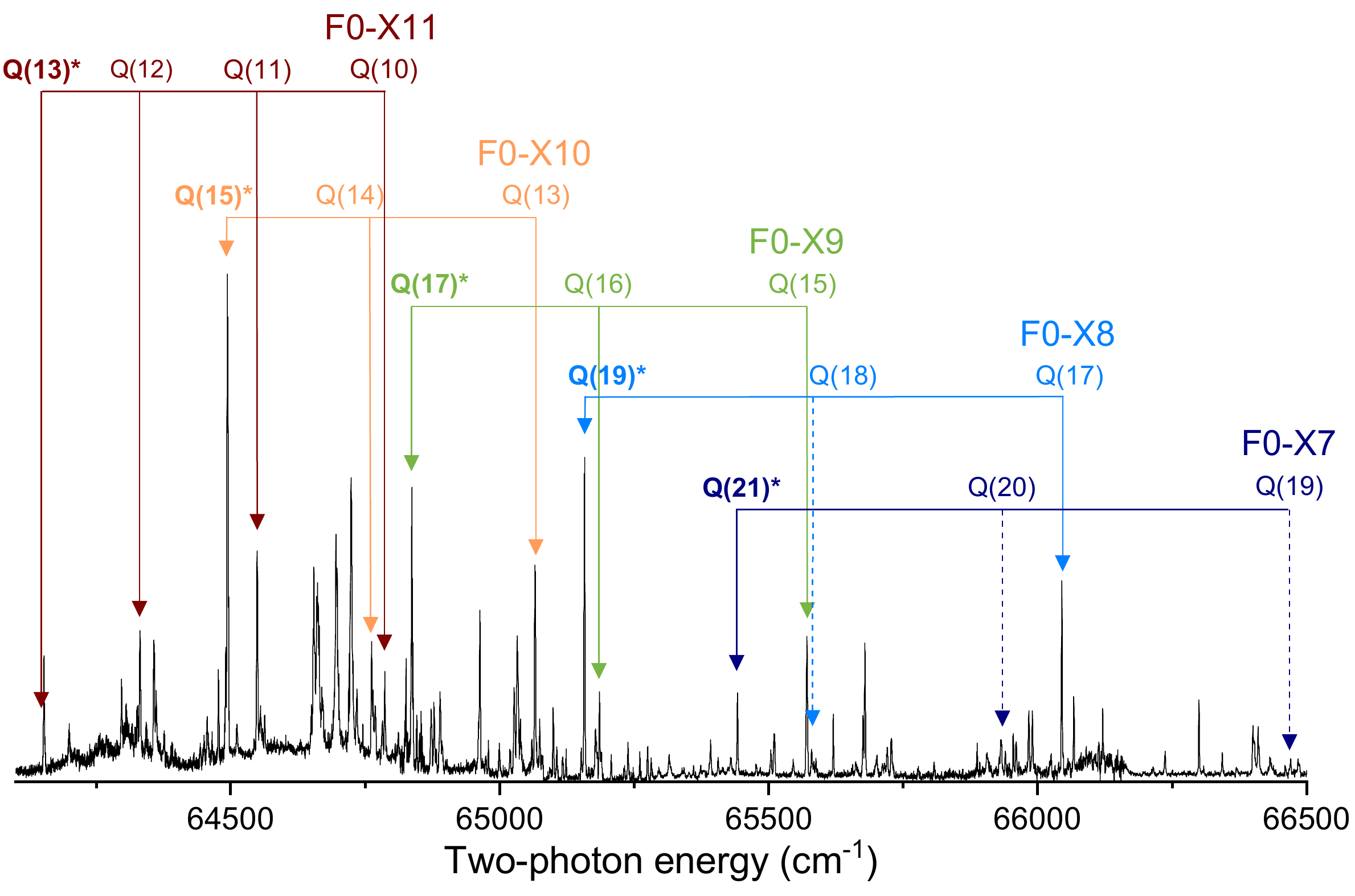}
\includegraphics[width=0.29\linewidth]{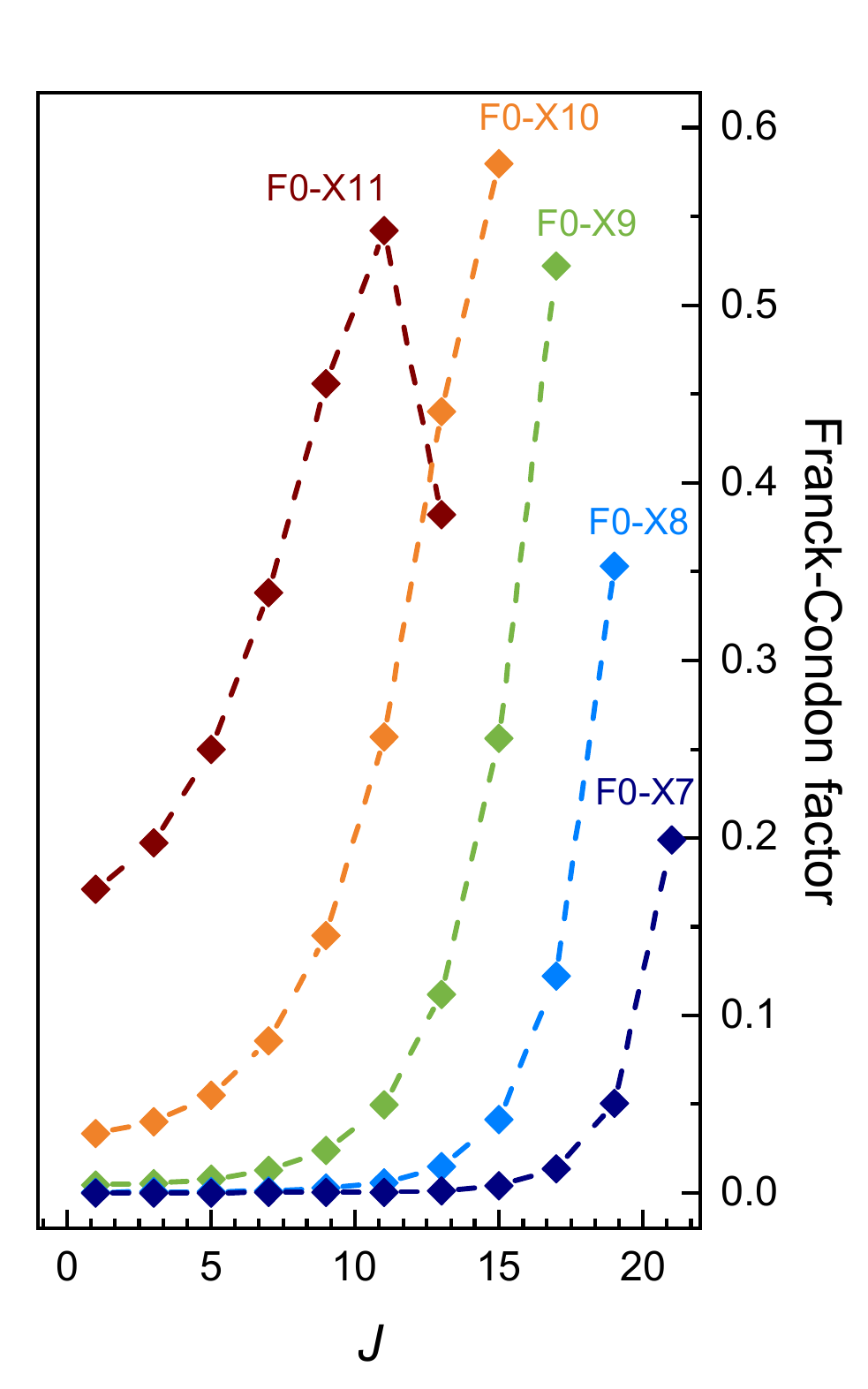}
\caption{\label{Overview}
(Left): Overview spectra recorded in a two-laser scheme with two-photon UV-photolysis of H$_2$S, followed by 2+1 REMPI on F-X($v',v''$) bands with a UV-tunable frequency-doubled dye laser. Transitions are labeled with quantum numbers of the ground level.
Excitations from quasi bound resonances are indicated with asterisk (*). The dashed arrows represent the calculated positions of unobserved transitions. (Right): Calculated Franck-Condon factors for the bands F-X(0,v'') as a function of $J$.
}
\end{center}
\end{figure*}

\subsection{The \F\ state}
\label{sec:theoryF}
To predict transition frequencies for the F0-X transitions, term values of the \EF\ electronic state were calculated using Born-Oppenheimer \cite{Silkowski2021}, adiabatic \cite{Wolniewicz1994} and relativistic \cite{Wolniewicz1998} potential energy curves.
Leading order radiative corrections were taken into account by using the corresponding curves of the hydrogen molecular ion \cite{Korobov2018}.
Nonadiabatic energies for the F0 states are reported for $J=0-5$ in Ref.~\cite{Yu1994} and were obtained from a coupled-equations calculation including several \textit{gerade} states.
In the spirit of Ref.~\cite{Wolniewicz1986a} the energy difference between the reported nonadiabatic and the current adiabatic term values are parametrized as $\Delta T_\text{NA}=a + bJ(J+1)$, relating $a$ to homogeneous and $b$ to heterogeneous interactions.
The found parameters $a=0.67$~cm$^{-1}$ and $b=-0.009$~cm$^{-1}$ are expected to predict F0 term values with an accuracy of around 1~cm$^{-1}$.
These renewed computations for the F0($J$) levels energies were performed since values for the highest $J$-levels were lacking in \cite{Dickenson2012a}. Based on a comparison with experimental values for F0($J$) at low $J$ \cite{Bailly2010} the present computations are shown to be more accurate.
Values for F0($J$), for the $J$-values relevant to the present study, are listed in Table~\ref{tab:transition}.

\section{Detection, identification and precision measurement of the quasi-bound resonances}
\label{detect}

The product distribution of rovibrationally excited states in H$_2$, upon two-photon photolysis of H$_2$S, was first measured in an overview scan in the wavelength range 300-310 nm and displayed in Fig.~\ref{Overview}.
The overview spectrum shows many lines in ($v',v''$) vibrational bands in the \F-\X\  system, detected via 2+1 resonance-enhanced multiphoton ionization.
Progressions of O($\Delta J=-2$), Q($\Delta J=0$) and S($\Delta J=2$) rotational branches are overlaid, thus forming a dense spectrum.

The assignment of the F0-X transitions originating in quasi-bound resonances is based on a comparison with computed level separations between  \X\ and \F\ levels as presented in section~\ref{theory}.
To obtain the excitation energies of H$_2^*$ above \X($v=0,J=0$) the values for the binding energies of X($v,J$) are augmented with the value for the dissociation energy of H$_2$, for which the most recent experimental value was taken, $D_0=36118.069605(31)$~cm$^{-1}$ \cite{Beyer2019}, which is in excellent agreement with the theoretical value of $D_0 = 36 118.069 632\,(26)$ \wn~\cite{Puchalski2019b}.
From a combination of these values, included in Table~\ref{tab:transition}, a prediction can be made for the F0-X transitions probing the quasi-bound resonances.
Although the predictions are systematically off by $\sim1$ \wn\ from measurement, they can be considered proof for the assignment of the H$_2$($v,J$)$^*$ levels.
It is noted that the found deviations from experiment simply reflect the inaccuracy in the ab initio calculations for F0($J$) levels, which is by itself considered very good for electronically excited states in H$_2$.
The transitions are assigned in Fig.~\ref{Overview}, where the transitions originating in quasi-bound resonances are marked with an asterisk (*).

\begin{figure*}
\begin{center}
\includegraphics[width=\linewidth]{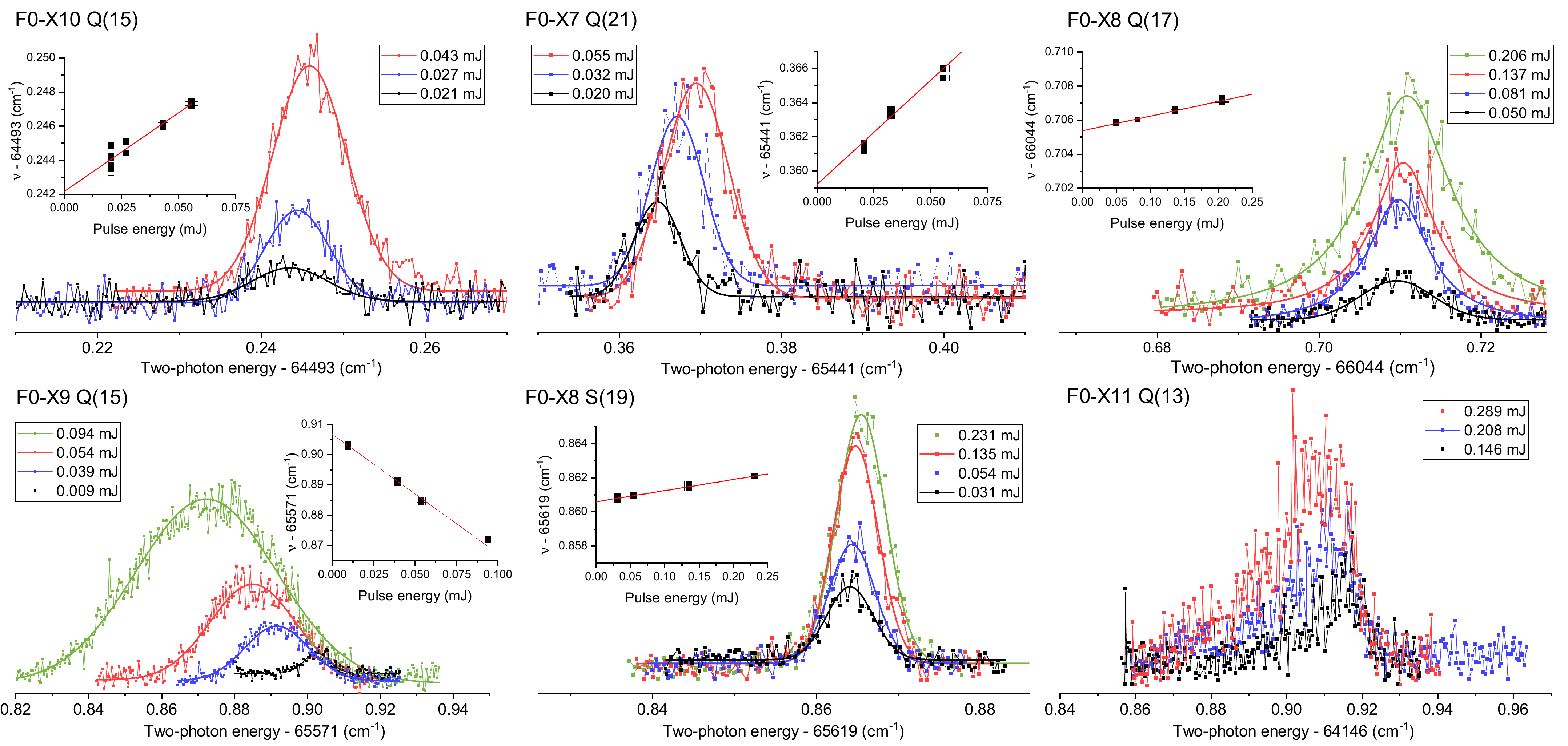}
\caption{\label{QB-spectra}
Excitation spectra of two photon transitions in the F-X system probing a number of quasi-bound resonances H$_2^*$, including from the X(11,13)$^*$ short-lived one, and from the bound state  X(9,15).
The insets show the AC-Stark extrapolation from measurements at varying pulse energies to zero power levels.}
\end{center}
\end{figure*}

The overview spectrum of Fig.~\ref{Overview} shows  that the two-photon excitations originating in the H$_2^*$ quasi-bound resonances are markedly more intense than the excitations for the bound resonances at lower rotational quantum number.
This apparent deviation of the Franck-Condon principle is a direct consequence of the shift of wave function density of the quasi-bound levels to large internuclear separation (see also Fig.~\ref{ex_scheme}).
A computation of $J$-dependent Franck-Condon factors for the F0-X transitions, results of which are displayed in Fig.~\ref{Overview}, show this exquisite behavior. The Franck-Condon factors were calculated using the adiabatic wave functions and using box-normalization for the quasi-bound states.
This provides further proof for an unambiguous assignment of the quasi-bound states detected experimentally.

In the experimental configuration with three UV-lasers non-overlapping in time, four quasi-bound resonances, X(7,21)$^*$, X(8,19)$^*$, X(9,17)$^*$ and X(10,15)$^*$, are probed in a  precision scheme, with the narrowband PDA-laser detecting the Doppler-free \F-\X\ two-photon  transition.
In Fig.~\ref{QB-spectra} several examples of the precision frequency measurements are presented, exciting quasi-bound resonances, further referred to as H$_2^*$, into the electronically excited F0 state, with a subsequent UV pulse for H$_2^+$ formation and detection.
An excitation spectrum from the bound state X(9,15) is also shown.
The short-lived X(11,13)$^*$ quasi-bound resonance, with a lifetime of 1.6~ns, was probed in an experimental configuration
with a 3~ns separation (and thus in part temporally overlapping) between dissociation laser and the (PDA)-probe laser, yielding a spectrum at low signal-to-noise ratio (see Fig.~\ref{QB-spectra}).

The spectroscopic measurements are performed for varying intensities allowing for the assessment of the AC-Stark effect in extrapolation to zero field as shown in the insets.
All five quasi-bound states have been excited via a Q-transition ($\Delta J=0$), while some are also probed via S- or O-transitions.
The extrapolated zero-intensity values of the highly accurate transition frequencies are compiled in Table~\ref{tab:transition}.

For the higher intensities used, the AC-Stark effect is the determining factor in the spectral linewidth, while for the lowest intensities the laser bandwidth of the PDA significantly contributes. The narrowest spectrum obtained is about 200 MHz FWHM, close to the PDA linewidth (including effects of frequency doubling and two-photon excitation).
No broadening effect on the spectral line shape as a result of tunneling through the centrifugal barrier was found.
Resonances with X($v,J$)$^*$ = (7,21), (8,19) and (9,17) are predicted to have lifetimes in excess of 100 $\mu$s, while X(10,15)$^*$ lives for 2 $\mu$s~(see Table~\ref{tab:calc_res}).
These lifetimes (short with respect to the bound states) lead to increased natural linewidths, but this is still below the frequency resolution under the present experimental conditions.
In fact not even the short-lived X(11,13)$^*$ resonance exhibits a significant line broadening at the present resolution.

\begin{table*}
\renewcommand{\arraystretch}{1.4}
\setlength{\tabcolsep}{8pt}
\begin{threeparttable}
\caption{\label{tab:transition}
Measured frequencies for the two-photon F0-X transitions probing the quasi-bound levels H$_2^*$, with uncertainties indicated in parentheses. Also some transitions for bound H$_2$ levels are included, relevant for comparison.
Computed values from the present work for the F0($J$) levels are listed.
Term values of the quasi-bound resonances X($v,J$)$^*$ are obtained from the dissociation energies given in Table~\ref{tab:calc_res} and adding the most recent  theoretical values for the dissociation energy $D_0=36118.069605(31)$~cm$^{-1}$ \cite{Beyer2019}.
For the bound states values from H2SPECTRE~\cite{SPECTRE2019,Czachorowski2018} are taken.
From a combination of these a calculation of two-photon transitions frequencies F0-X can be made. Finally, in the last column the deviations between experimental and predicted frequencies are listed as $\Delta_{\rm Exp.-Calc.}$. All values in \wn.
}
\begin{tabular}{cclcccc}
X$(v,J)$	&  Transition & Exp.   &  F0($J$)$_{\rm calc}$  & X($v,J$)$^*_{\rm calc}$ & Calculated   & $\Delta_{\rm Exp.-Calc.}$ \\
\hline
 (7,21)* & Q(21) & 65\,441.3575 (9) &  102\,064.12   & 36\,624.0006  & 65\,440.12  & 1.24 \\
 (7,21)* & O(21) & 64\,979.430 (10)  &  101\,602.55   & 36\,624.0006  & 64\,978.55 & 0.88 \\
 (8,19)* & Q(19) & 65\,157.9413 (21) &  101\,602.55   & 36\,445.4988  & 65\,157.05  & 0.89 \\
 (8,19)* & S(19) & 65\,619.8599 (8) &  102\,064.12   & 36\,445.4988  & 65\,618.62  & 1.24 \\
 (8,19)* & O(19) & 64\,734.8098 (9)$^a$  &  101\,179.70   & 36\,445.4988  & 64\,734.20  & 0.61 \\
 (9,17)* & Q(17) & 64\,837.2974 (19) &  101\,179.70   & 36\,343.0107  & 64\,836.69  & 0.61 \\
 (9,17)* & O(17) & 64\,454.775 (10) &  100\,797.40   & 36\,343.0107  & 64\,454.39  & 0.39 \\
(10,15)* & O(15) & 64\,152.970 (20)$^b$ &  100\,457.32   & 36\,304.5238  & 64\,152.80  & 0.17 \\
(10,15)* & Q(15) & 64\,493.2404 (9) &  100\,797.40   & 36\,304.5238  & 64\,492.88  & 0.36 \\
(11,13)* & Q(13) & 64\,146.930 (20)$^b$ &  100\,457.32   & 36\,310.5641  & 64\,146.76  & 0.17 \\
\hline
 (8,17)  & Q(17) & 66\,044.7046 (9) &  101\,179.70   & 35\,135.6038  & 66\,044.10  & 0.61 \\
 (9,15)  & Q(15) & 65\,571.9063 (19) &  100\,797.40   & 35\,225.8567  & 65\,571.54  & 0.36 \\
 (9,15)  & S(15) & 65\,954.4505 (10) &  101\,179.70   & 35\,225.8567  & 65\,953.84  & 0.61 \\
\hline
\hline
\end{tabular}
\begin{tablenotes}
\footnotesize
\item $^a$ Measured both in DC-field free and in DC-field for prompt ion removal.
\item $^b$ Measured with 3 ns delay, hence temporal overlap, between dissociation laser and spectroscopy laser.
\end{tablenotes}
\end{threeparttable}
\end{table*}

The error budget for the frequency calibrations, presented in Table~\ref{tab:error}, contains a variety of contributions.
Minor contributions relate to the  calibration uncertainty of the cw-laser seeding the PDA, of some 2 MHz, while
a residual Doppler shift from misalignment of the counter-propagating beams is reduced by Sagnac interferometry to below 3 MHz uncertainty.
The chirp-induced frequency correction accounts for another 4.5 MHz uncertainty.
The AC-Stark effect yields the largest contribution to the error budget.
It is addressed by performing systematic measurements resulting in the AC-Stark slopes as indicated in Fig. \ref{QB-spectra}.
The uncertainty associated with AC-Stark results from the extrapolation to zero-power levels and depends on the obtained signal-to-noise ratio for individual lines.
In order to reduce the contributions of the AC-Stark effect various campaigns of remeasurement of the Stark-slopes were carried out, thus turning the systematic effect into a statistical distribution of results.
For some weak transitions, for example F0-X11 Q(13) and F0-X10 O(15), this could not be done effectively, and only measurements at high laser power were performed leading to larger uncertainties.
For each resonance targeted and listed in Table~\ref{tab:transition} the  resulting final experimental uncertainty was deduced from the statistical and systematic contributions and by taking them in quadrature.
This leads to a variety of resulting uncertainties, mainly associated with the number of measurement campaigns for each line and the obtainable signal-to-noise ratio.
For the Q-branch lines an optimum accuracy was obtained at the level of 0.001 \wn, corresponding to 30 MHz.
It is noted that the overall measurement uncertainty has been  improved in comparison with previous work~\cite{Lai2021}; this is related to increased statistical averaging and stronger signal strength resulting from larger FC-factors for the measured transitions.

\begin{table}
\renewcommand{\arraystretch}{1.2}
\caption{\label{tab:error}
Uncertainty budget for the measurements of two-photon F0-X transitions. The uncertainty for the AC-Stark extrapolations is estimated for individual transitions, where the total uncertainties are listed in Table~\ref{tab:transition}.
}
\begin{tabular}{lc}
Contribution & Uncertainty (MHz) \\
\hline
Lineshape fitting & 15 \\
Frequency calibration & 9 \\
CW-pulse offset (chirp)  & 18 \\
Residual Doppler effect & 3 \\
DC-Stark effect & < 1\\
\hline
Subtotal (exl. AC-Stark) & 25\\
AC-Stark effect &  $3 - 60$\\
\end{tabular}
\end{table}

\begin{figure*}
\begin{center}
\includegraphics[width=0.75\linewidth]{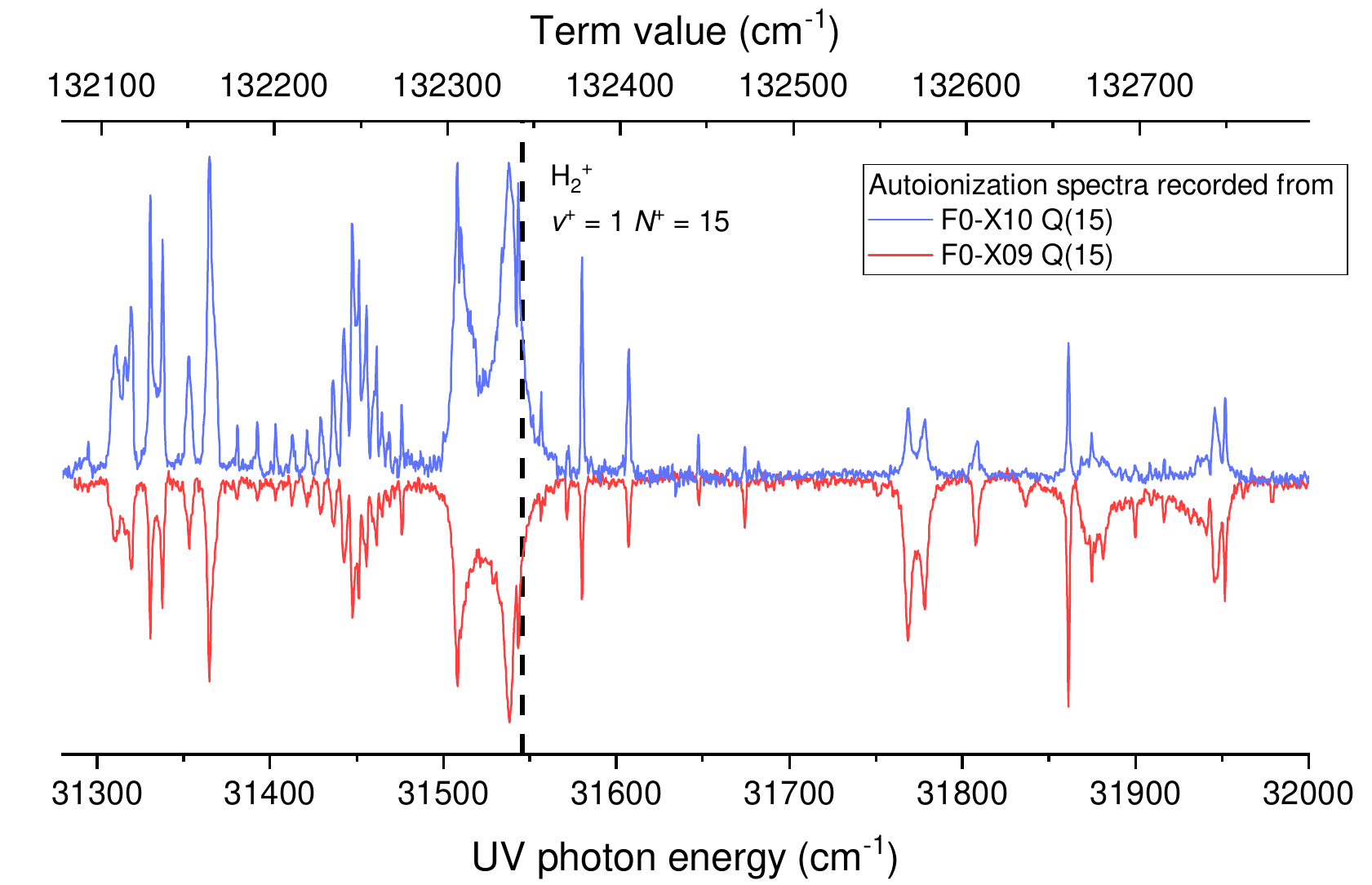}
\caption{\label{Auto15}
Two recordings of autoionization spectra after probing different two-photon transitions in the F-X system, one from the bound X(9,15) state and one from the quasi-bound X(10,15)$^*$ resonance.
The similarity of the resonances in both spectra and the observation of a Rydberg series converging to the X$^+$($v^+=1$,$N^+=15$) limit in H$_2^+$ proves that a $J=15$ state is probed at the intermediate level.}
\end{center}
\end{figure*}

During the recording of the F0-X two-photon transitions by tuning the narrowband PDA-laser (see Fig.~\ref{QB-spectra}) the third UV-laser  is parked on an arbitrary autoionization resonance for optimum H$_2^+$ signal generation.
Alternatively, when signal is found, either on bound states or on quasi-bound resonances, the wavelength of the third laser may be scanned while keeping the spectroscopy laser fixed.
By this means the energy range in H$_2$ above $130\,000$ \wn\ is probed relying on autoionization.
In Fig.~\ref{Auto15} autoionization spectra are displayed that were recorded after setting the spectroscopy laser on the two-photon transitions from X($v=9,J=15$) bound and X($v=10,J=15$)$^*$ quasi-bound states.
The autoionization spectra are plotted against the term value corresponding to the total excitation energy in the H$_2$ molecule.
The one-to-one correspondence between the autoionization resonances proofs that these spectra are taken from an intermediate state with the same angular momentum $J$, thus proving that the quasi-bound resonance has $J=15$.
Note that the small differences in the intensities are caused by the differences in (vibrational) wave function overlap for $v=9$ and $v=10$ levels and the ionic states.
The fact that an autoionizing Rydberg series is observed converging to the H$_2^+$($v^+=1,N^+=15$) provides further evidence that the intermediate state is $J=15$.
Such autoionization spectra were recorded for all  quasi-bound resonances, therewith verifying their $J$-quantum numbers.
Apart from providing additional proof on the assignment of resonances, this method gives access to H$_2$ Rydberg states with unprecedentedly high rotational angular momentum of the H$_2^+$ ion core, to be explored in future.

\section{Discussion}

Five quasi-bound resonances were detected in a precision experiment. These resonances had previously been observed as terminal levels in emission via the Lyman and Werner bands in studies by Dabrowski \cite{Dabrowski1984} and  by Roncin and Launay \cite{Roncin1994} at an accuracy of 0.05 \wn.
The precision of the present Doppler-free laser excitation study is much improved, by more than an order of magnitude.
The novelty of the present work derives not only from the improved accuracy, but also on the fact that the centrifugally bound H$_2^*$ resonances could be produced as a transition state in a photolysis process and thereupon detected via 2+1' resonance-enhanced multi-photon ionization before tunneling and separating into two H($1s$) atoms.

\subsection{Test of the H$_2$ potential}
\label{Pot-test}

\begin{figure}
\begin{center}
\includegraphics[width=\linewidth]{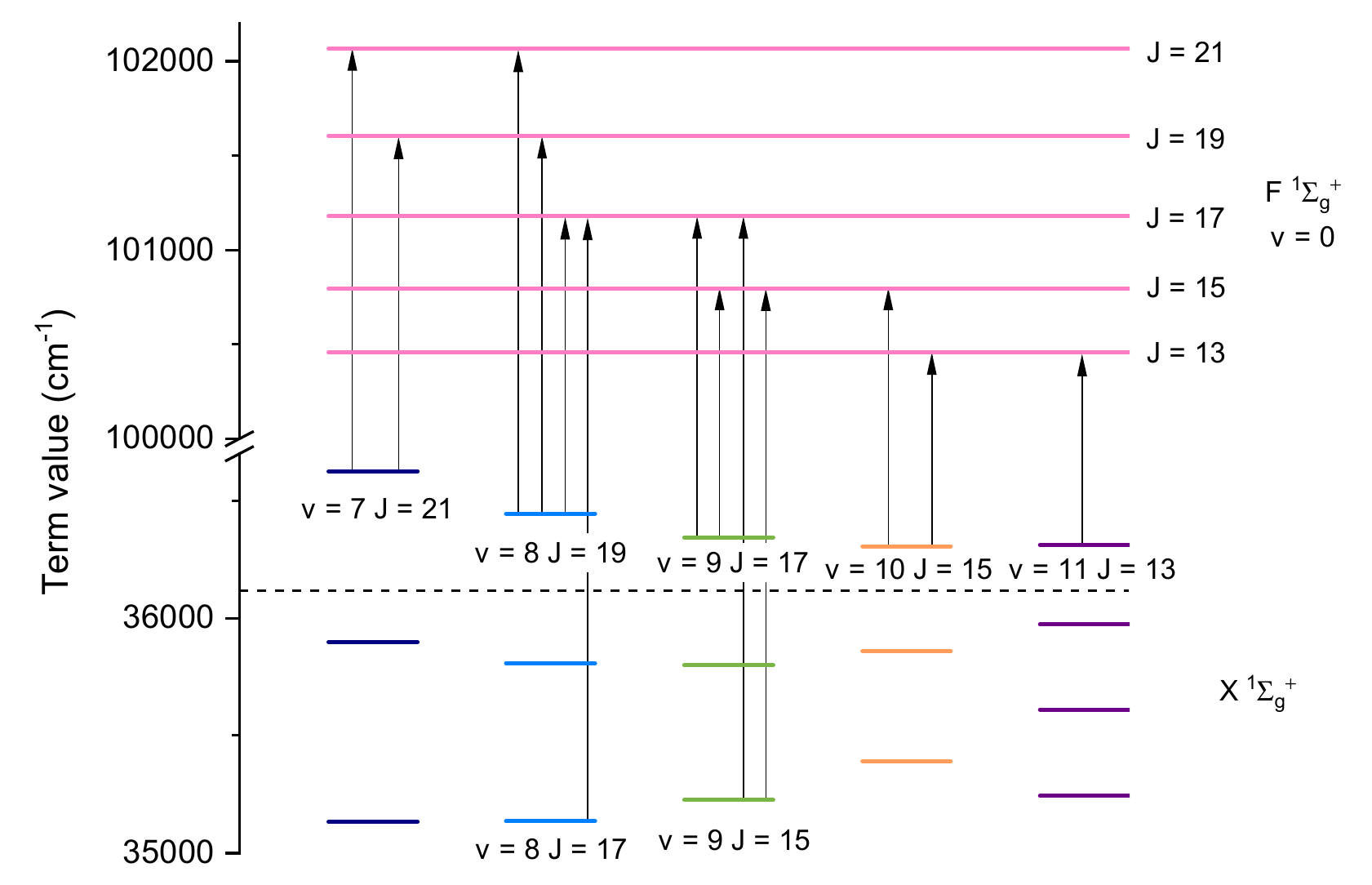}
\caption{\label{comb_diff_graph}
Level diagram of X($v,J$) and F0($J$) rotational levels with the H$_2$(\X) dissociation limit given as a dashed line. Measured transitions, plotted as (black) arrows, connect all the observed quasi-bound resonance including measurements of weak S- and O- transitions. The energy splittings are determined from the combination differences and listed in Table~\ref{tab:comb_diff}.}
\end{center}
\end{figure}

Although the quasi-bound resonances were probed in a precision experiment the detection pathway via the \F\ outer well state does not provide a means to determine the (negative) binding energy of the resonances in a direct manner.
Hence, the accurate first principles computations of H$_2^*$(X), at an accuracy of 0.003 - 0.004 \wn\ (cf. Table~\ref{tab:calc_res}), cannot directly be compared with observation.
In order to make a comparison possible, combination differences between resonances of bound and quasi-bound nature all lying near threshold, have been determined via the level diagram plotted in Fig.~\ref{comb_diff_graph}.
The  energy of all resonances is deduced relative to the X(10,15)$^*$ level and the experimental uncertainties are included via error propagation.
Theoretical values for these combination differences are
computed and a comparison with experiment is made in Table~\ref{tab:comb_diff}.
The uncertainty of the higher order terms in the NAPT approach are related to an interaction with distant electronically excited states.
For this reason, the uncertainties for the individual X($v$,$J$) levels  are not independent of each other, so that we estimated the accuracy of the calculated intervals by multiplying the non-adiabatic correction of the interval with $1/\mu_\text{n}$.
Its outcome demonstrates that the deviations between experimental and computed combination differences are well within 0.001 \wn\ (30~MHz).

\begin{table}[t]
\renewcommand{\arraystretch}{1.3}
\caption{Combination differences between H$_2^*$ quasi-bound resonances, and some bound resonances. Level energies are plotted with respect to the X(10,15)$^*$ resonance. A comparison is made between experimental determination and computed values, while in the last column deviations are given. Note that these discrepancies relate to the inaccuracy in the calculation of F0($J$) levels, and not to inaccuracies in the bound and quasi-bound resonances on the ground state potential.
All values in \wn.
\label{tab:comb_diff}}
\begin{tabular}{ld{8}d{8}d{8}}
X$(v,J)$ & \multicolumn{1}{c}{$T_\text{rel}^\text{exp}$} & \multicolumn{1}{c}{$T_\text{rel}^\text{calc}$}   & \multicolumn{1}{c}{$\Delta T$}   \\
\hline
(7,21)$^*$  &   319.4772\,(28)    &   319.4768\,(9)    &   0.0004\,(29) \\
(8,17)   &   -1168.9200\,(25)  &   -1168.9200\,(12)  &   0.0000\,(28) \\
(8,19)$^*$  &   140.9748\,(25)    &   140.9750\,(7)    &   -0.0002\,(26) \\
(9,15)  &   -1078.6659\,(21)  &   -1078.6671\,(11)  &   0.0012\,(24) \\
(9,17)$^*$  &   38.4872\,(30)     &   38.4869\,(4)      &   0.0003\,(30) \\
(10,15)$^*$ &   \multicolumn{1}{c}{0}               &   \multicolumn{1}{c}{0}               &   \multicolumn{1}{c}{0}\\
(11,13)$^*$ &   6.041\,(28)      &   6.0404\,(6)       &   0.001\,(28) \\
\hline
\end{tabular}
\end{table}

This excellent agreement constitutes a test of the high accuracy of the calculated potential energy curve of H$_2$, including all adiabatic, non-adiabatic, relativistic and QED contributions as discussed in section~\ref{theory}.
The unprecedented experimental accuracy for highly excited vibrational states verifies the computations in the NAPT-framework \cite{Czachorowski2018} for the H$_2$ potential specifically at large internuclear distances.

This comes in addition to the tests of the theoretical framework near the bottom part of the H$_2$ potential.
The rotationless fundamental vibrational splitting ($\Delta v'=1-0$) in H$_2$ was measured to an accuracy of 5 MHz in a molecular beam experiment \cite{Dickenson2013}.
This experimental value is in agreement within a single standard deviation, both with NAPT (uncertainty of 27 MHz) and with the more precise 4-particle non-Born-Oppenheimer calculation (at 9 MHz uncertainty).
The S(3) line in the second overtone ($\Delta v'=3-0$) was measured to high accuracy in Doppler-broadened cavity-ring down spectroscopy \cite{Cheng2012,Hu2012}.
A reanalysis of its collisional line shape, yielding an accuracy of 6 MHz \cite{Wcislo2016}, is again found in agreement with theory.
The measurement of the first overtone ($\Delta v'=2-0$) reached a lower accuracy of 30 MHz \cite{Kassi2014}, also in agreement with theory. The Q($J=1,3,5,7$) level splittings for $\Delta v'=11-13$ were determined with an accuracy between 84 and 93~MHz by using combination differences of F-X transitions, also agreeing with NAPT calculations~\cite{Lai2021}.
Measurements of the dissociation energy ($D_0$) of H$_2$ also probe the bottom part of the potential.
The most recent experimental determinations of $D_0$(H$_2$) \cite{Cheng2018,Beyer2019,Holsch2019} are found to be in agreement with theory \cite{Puchalski2019b} at the level of 1 MHz.

\begin{figure}
\begin{center}
\includegraphics[width=1\linewidth]{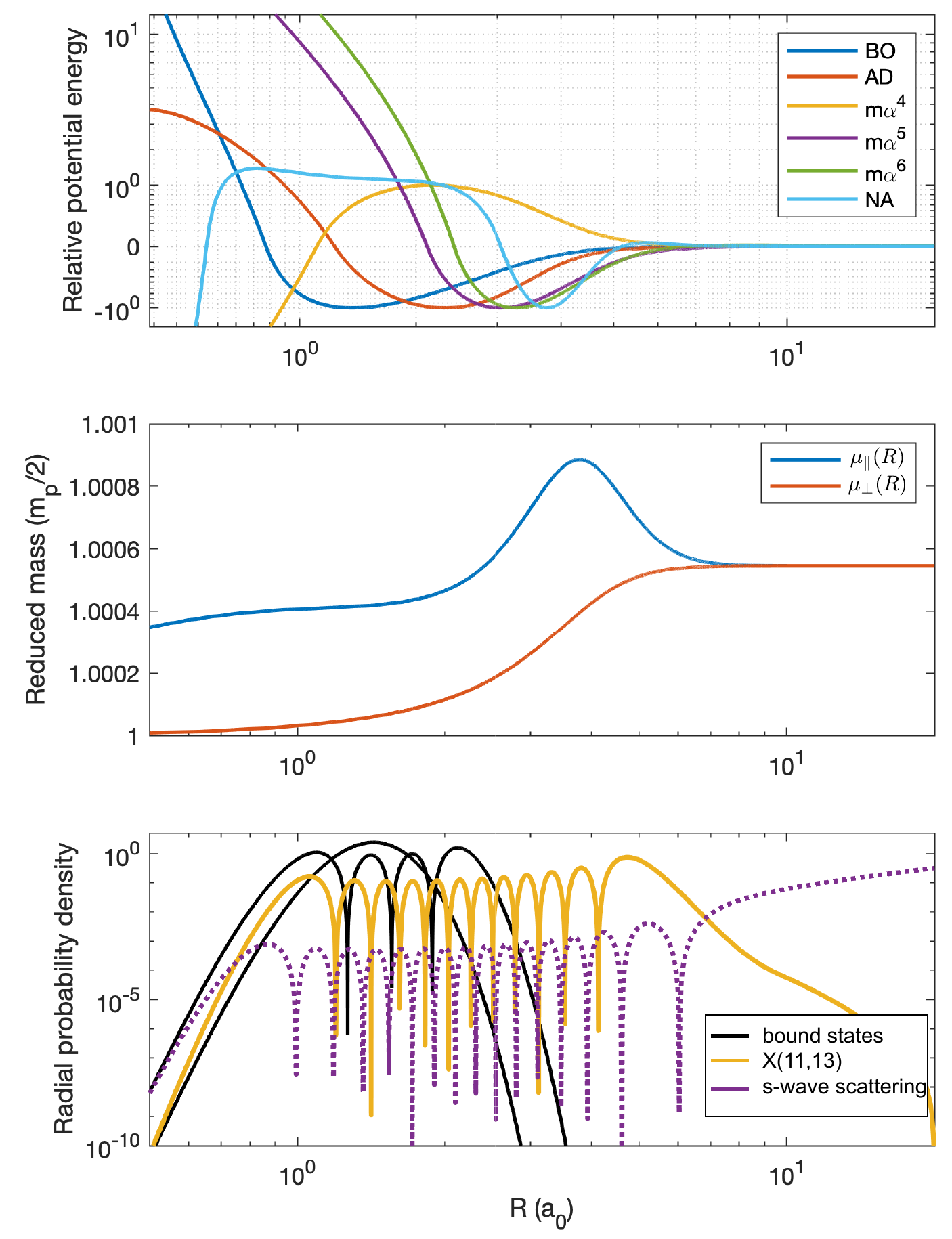}
\caption{\label{fig:PESrange}
Upper panel: Range of the BO potential energy curve and the various corrections normalized to $+1$ or $-1$ around the mean internuclear distance for better comparison.
Middle panel: $R$-dependent vibrational and rotational reduced masses, $\mu_{\|}(R)$ and $\mu_{\perp}(R)$, used to take non-adiabatic effects into account within the framework of NAPT.
Lower panel: Radial probability density of the $v=0,3$ bound states (black solid line) and the X(11,13) quasi-bound state (colored solid line). The squared $s$-wave zero-energy wave function is shown for comparison (dotted line, not to scale).
}
\end{center}
\end{figure}

The combination of these experimental data allows the first specific test of the H$_2$ potential energy curves and the $R$-dependent reduced masses at the 90~MHz level over the entire relevant range of internuclear distances, as illustrated in Fig.~\ref{fig:PESrange}.
The upper panel of Fig.~\ref{fig:PESrange} displays the BO, adiabatic, non-adiabatic, relativistic and QED potential energy curves normalized to $\pm1$ around $R\approx 2.5\,a_0$ for better comparison. The absolute values of the displayed BO, adiabatic and $m\alpha^4$ relativistic potential energy curves normalized to $\pm1$ amount to ${\sim}38293$~\wn, ${\sim}18$~\wn\ and ${\sim}0.9$~\wn, respectively.
Similarly, one finds 1.6~GHz, 8~MHz and 0.66~GHz for the $m\alpha^5$, $m\alpha^6$ and the non-adiabatic correction. The middle panel shows the $R$-dependent reduced vibrational and rotational masses, $\mu_{\|}(R)$ and $\mu_{\perp}(R)$, varying smoothly from the nuclear reduced mass at $R\to0$, to the atomic reduced mass for $R\to\infty$.
The lower panel displays the squared vibrational wavefunctions of the bound states for $v=0$ and 3 (black solid line), and the observed X(11,13) quasi-bound state (colored solid line).
It can be seen that the previous experimental studies with $v<3$ tested the potential mainly for $R$ below 2.5~$a_0$.
For the observed state X(3,2), 99\% of the radial probability density is within the interval $[0.9, 2.5]a_0$, whereas the interval $[1, 6.6]a_0$ is presently probed by the $X(11,13)$ state.
At $6.6a_0$ the $R$-dependent reduced rotational and vibrational masses only deviate by 0.0003\% and 0.002\% from the asymptotic atomic reduced mass, respectively. The various parts of the potential reached their respective asymptotic value to within 2\% or less, which amounts to 76~\wn, 0.09~\wn\ and 0.01~\wn\ for the BO, adiabatic and relativistic potential and less than 42~MHz for the remaining contributions.

\begin{figure*}
\begin{center}
\includegraphics[width=1\linewidth]{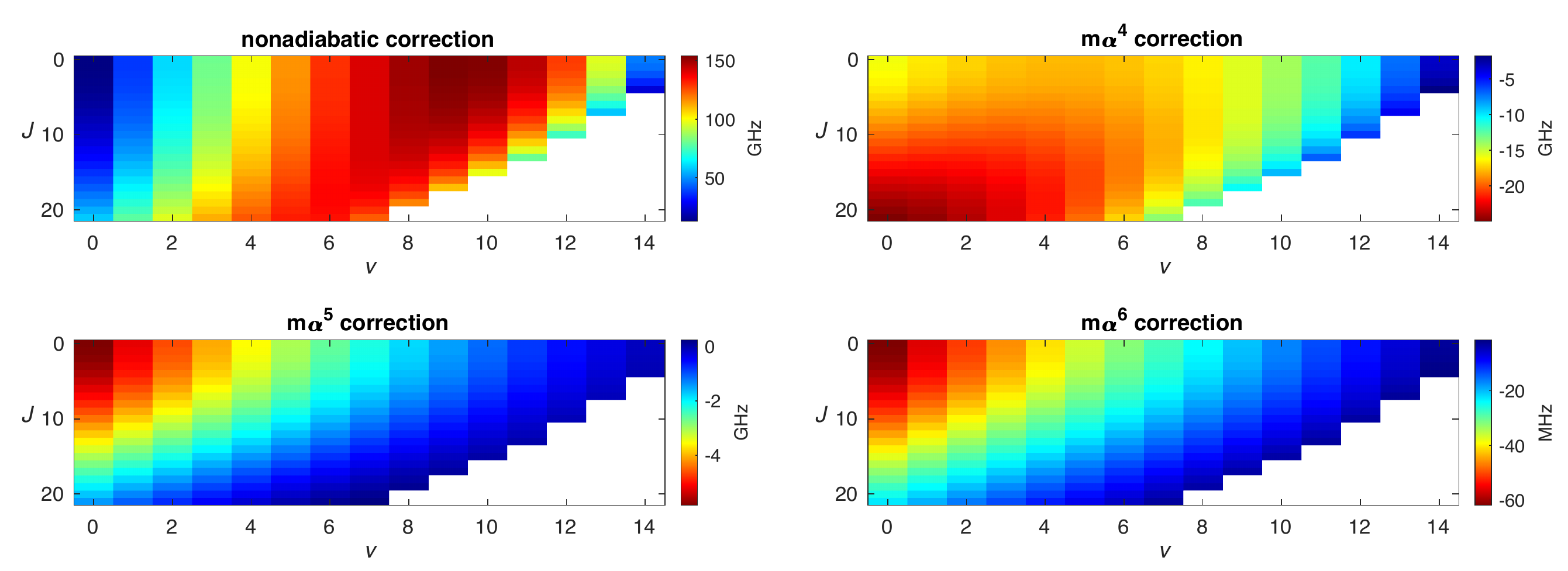}
\caption{\label{fig:corr-heatmap}
Nonadiabatic, $m\alpha^4$, $m\alpha^5$ and $m\alpha^6$ corrections for the bound states and the long-living quasi-bound states of H$_2$ for $J=0-21$. Note that the uncertainties for the separate contributions scale with the actual values plotted here (see text).
}
\end{center}
\end{figure*}

In view of confronting experimental values for testing theoretical approaches it is illustrative to consider the numerical contributions of the nonadiabatic, relativistic and orders of QED corrections as included in the computations of section~\ref{theory}.
In Fig.~\ref{fig:corr-heatmap} the various contributions to the total energy are graphically displayed in a ($v,J$) color plot.
While  $m\alpha^n$ ($n=4-6$) for relativistic and QED-contributions is smallest for the highest $v$-levels, the non-adiabatic correction has the largest contribution for vibrational levels $v=7-11$, hence for the quasi-bound states probed in the current experiment.
In the same manner Fig.~\ref{fig:corr-heatmap} illustrates that the $m\alpha^4$ relativistic term is best tested through precision experiments on levels with low-$v$ and high-$J$ as was done previously~\cite{Salumbides2011}.

It was long understood that the magnitude of the nonadiabatic corrections is related to the vibrational kinetic energy, which reaches its maximum for $v=9$.
That turns the present precision measurement of quasi-bound resonances into a specific test of computations of the non-adiabatic correction term in the computation of H$_2$ level energies.
It should be noted that in Fig.~\ref{fig:corr-heatmap} the values for the energy contributions are plotted, and that their uncertainties scale with the absolute values \cite{Komasa2019} ($1/\mu_\text{n}$, $<2\times10^{-4}$, $5\times10^{-4}$, $3\times10^{-3}$, respectively).

The dissociation energy sets the most stringent test of the potential energy curves, because it probes the internuclear distances where the corrections are largest,
as can be seen by comparing the top and lower panel of Fig.~\ref{fig:PESrange}.
The quasi-bound states, however, allow to test the potentials over a much wider range, while being particularly sensitive to the non-adiabatic effects.
By scaling the individual potential energy surfaces and repeating the calculation of the dissociation energy $D_0$ (and quasi-bound intervals, given in brackets), upper limits are found to be: BO: $10^{-9}$ ($5\times10^{-7}$), AD: $5\times10^{-6}$ ($8\times10^{-4}$), $m\alpha^4$: $6\times10^{-4}$ ($2\times10^{-2}$), $m\alpha^5$: $2\times10^{-4}$ ($4\times10^{-1}$), $m\alpha^6$: $2\times10^{-2}$ ($3\times10^{1}$) and NA: $1\times10^{-3}$ (1), respectively.
Using the quasi-bound intervals we also constrained a relative error of $W_\|$ and $W_\perp$ to 0.05 and 0.002, respectively.

\subsection{H + H scattering}
\label{HHscatter}

The H + H scattering process is of fundamental interest in physics, playing a role in the formation of molecular hydrogen in the universe \cite{Forrey2016}, frequency shifts in the hydrogen maser \cite{Wittke1956}, the magnitude of pressure shifts in the 1S-2S transition of atomic hydrogen for a determination of the Rydberg constant \cite{Jentschura2019,Matveev2019}, and for the formation of a Bose-Einstein condensate in hydrogen \cite{Fried1998}.
Because of the light atomic mass, non-adiabatic effects are especially important in this collision process and the question about the correct treatment of these was vividly discussed in the literature \cite{Jamieson2010} (and references therein).

The scattering process can be described by a single parameter at low temperature, the scattering length $a$ given by
\begin{equation}
    a = -\lim_{k\to 0} \frac{\tan \eta_{J=0} }{k}.
\end{equation}
Whereas previously reported scattering lengths obtained by different authors within the BO and adiabatic approximations were found to agree, the values for the non-adiabatic scattering length varied between 0.3006~$a_0$ and 0.564~$a_0$ (see Table~3 in Ref.~\cite{Jamieson2010} and references therein), depending on the used reduced masses and effective correction potentials employed to account for the non-adiabatic interactions.
Moreover, Wolniewicz attempted a direct solution of the coupled equations \cite{Wolniewicz2003a} but encountered elementary difficulties because of spurious nonadiabatic couplings for $R\to\infty$ related to the choice of coordinates used to describe molecular states, not suited for the description of the free atoms in the asymptotic region \cite{Belyaev2010}.

We choose here the NAPT approach that was experimentally verified with MHz precision up to $R\sim 6.6\,a_0$, noting that at this distance the $R$-dependent masses and all potentials but the BO, adiabatic and relativistic potentials reached their asymptotic value (see section~\ref{Pot-test}).
The squared s-wave scattering wave function is displayed for comparison in the lower panel of Fig.~\ref{fig:PESrange}.
A comparison with the radial densities for the bound and quasi-bound states indicates that the whole range of internuclear distances with a significant potential energy term are probed.
In addition, as can be seen from the middle panel of Fig.~\ref{fig:PESrange}, the hump in the vibrational reduced mass $\mu_{\|}(R)$ is completely located within the interval probed by the resonances, which allows a verification of the main part of the non-adiabatic correction to the scattering length.
This allows for the first time to calculate scattering lengths employing an experimentally verified approach for taking the non-adiabatic interactions into account.
Using the techniques presented in Section \ref{sec:theoryX} we integrate the wavefunction outwards at $k_0=1\times10^{-7}$~a.u. to $R_\text{max}=500~a_0$ and obtain a scattering length for various level of approximation:
\begin{equation}
a=    \begin{cases}
      0.5699~a_0 & \text{BO}\\
      0.4160~a_0 & \text{AD}\\
      0.2572~a_0 & \text{NA} \\
      0.2735~a_0 & \text{NA},m\alpha^4,m\alpha^5,m\alpha^6.
    \end{cases}
\end{equation}

Using the dispersion coefficients for the BO \cite{Yan1996} and adiabatic potentials \cite{Przybytek2012} (the exchange interaction is negligible at this distance) we confirm that the asymptotic scattering length is obtained by using the extrapolation procedure given in \cite{Szmytkowski1995}. The value obtained for the adiabatic scattering length agrees with the value given by Wolniewicz \cite{Wolniewicz2003a}. It is also interesting to note, that the approximate nonadiabatic scattering length obtained using the atomic reduced mass is $0.2651~a_0$, deviating by only 3\% from the value obtained using NAPT. The deviation to previously reported values obtained using the same approximation is related to the improved BO and adiabatic potential energy curves \cite{Jamieson1998}.

The obtained scattering lengths are verified by comparing to \cite{Wu1962}
\begin{equation}\label{eq:scatleng_var}
    a = \frac{2\mu_n}{k^2} \int_0^\infty \sin(kR) \mathcal{V}(R)\chi(R;k=k_0) d\,R,
\end{equation}
and we find agreement at the 0.02\% level for the BO and adiabatic level of approximation.
As was found in Ref.~\cite{Jamieson1998}, \eqref{eq:scatleng_var} can also be used to estimate changes of the scattering length $\delta a$ caused by a change of the potential $\delta\mathcal{V}(R)$, by replacing $\mathcal{V}(R)\to\delta\mathcal{V}(R)$ and $\sin(kR)\to\chi(R;k=k_0)$.
We find $\delta a^{m\alpha^4} = 0.0156~a_0$, $\delta a^{m\alpha^5} = -0.0007~a_0$ and $\delta a^{m\alpha^6} = 0.0014~a_0$, respectively.

The experimental verification of the NAPT approach for the shape resonances can be used to determine an uncertainty of the scattering length. We studied the change of the scattering length when using the above-mentioned relative uncertainties for the individual parts of the potential or the $R$-dependent reduced masses.
This allows to attribute the experimental uncertainty of the shape-resonance intervals (${\sim}90$~MHz) to only one part of the nuclear Schr\"odinger equation. It is found that a change of the vibrational reduced mass $\mu_{\|}$ has the largest effect, resulting in an experimentally verified singlet scattering length of
\begin{equation}
a = 0.2735^{39}_{31}~a_0.
\end{equation}


\section{Conclusion}

In the present study five quasi-bound resonances of the H$_2$ molecule are produced in the two-photon UV photolysis of H$_2$S, where four of those persist as long-lived transition states.
Proof of their production and detection is provided by comparing experimental two-photon transition frequencies to computed level splittings between \F\ and \X\ levels.
Computation of rotational-state dependent Franck-Condon factors, compared with the observation of enhanced intensities, provides further verification of the assignments.
Also the step-wise excitation from quasi-bound states into the continuum provides an angular momentum label as supporting evidence.

Highly accurate calculations of \X($v$,$J$) level energies are performed, for which an existing framework of non-adiabatic perturbation theory (NAPT) \cite{Czachorowski2018,SPECTRE2019} is extended into the region above the dissociation threshold.
The present precision measurement allows for a test of the H$_2$ potential energy curve, comprising a Born-Oppenheimer potential, adiabatic, non-adiabatic and relativistic corrections, as well as QED-corrections up to order $m\alpha^6$.
The H$_2$ potential is tested now over a wide range of internuclear separations and  energies, by comparison of computed level energies to a set of data on infrared transitions, dissociation energy and quasi-bound resonances at MHz accuracy.
The precision measurement of the quasi-bound resonances specifically probes and tests the non-adiabatic correction to the H$_2$ potential.
This well-tested H$_2$ potential can be applied in a computation of the scattering length, resolving a decade old disagreement in the determination of the non-adiabatic singlet s-wave scattering length of the H(1s) + H(1s) collision.

\section*{Acknowledgement}

The authors thank Prof. Frederic Merkt (ETH Z\"urich) for discussions motivating this work. This work is financially supported by the European Research Council through an ERC Advanced grant (No: 670168). MB acknowledges NWO for a VENI grant (VI.Veni.202.140).

\end{document}